\documentclass[usenatbib]{mnras}

\usepackage{multirow}
\usepackage{rotating}
\usepackage{colortbl}
\usepackage{color}
\usepackage[fleqn]{amsmath}
\usepackage{amssymb}
\usepackage{amsfonts}
\usepackage{verbatim}
\usepackage{scalefnt}
\usepackage{overpic}
\usepackage[T1]{fontenc} 
\usepackage{aecompl} 
\usepackage{ulem}
\usepackage{cleveref}
\usepackage[utf8x]{inputenc}
\usepackage{threeparttable}
\usepackage{arydshln}
\usepackage{pdflscape}
\usepackage[referable]{threeparttablex}
\usepackage{footnote}
\usepackage{hyperref}
\usepackage{dirtytalk}
\usepackage{enumitem}
\usepackage{bm}
\usepackage[dvipsnames]{xcolor}

\newcommand{\JS}[1]{{ #1}}

\def\msun{\,{\rm M}_\odot}
\def\lsim{\mathrel{\rlap{\lower 3pt \hbox{$\sim$}} \raise 2.0pt \hbox{$<$}}}
\def\gsim{\mathrel{\rlap{\lower 3pt \hbox{$\sim$}} \raise 2.0pt \hbox{$>$}}}

\newcommand{\comments}[1]{} 

\title[Improved constraints from UFDGs on PBHs as DM]{Improved constraints from ultra-faint dwarf galaxies on primordial black holes as dark matter}

\author[J. Stegmann et al.]{Jakob Stegmann,$^{1,2}$\thanks{E-mail: StegmannJ@cardiff.ac.uk} Pedro~R. Capelo,$^{3}$ Elisa Bortolas$^{3}$ and Lucio Mayer$^{3}$\\
$^{1}$Department of Physics, ETH Zurich, Otto-Stern-Weg 1, 8093 Zurich, Switzerland\\
$^{2}$School of Physics and Astronomy, Cardiff University, Cardiff, CF24 3AA, United Kingdom\\
$^{3}$Center for Theoretical Astrophysics and Cosmology, Institute for Computational Science, University of Zurich,\\ 
Winterthurerstrasse 190, 8057 Zurich, Switzerland}

\date{Accepted 2020 January 16. Received 2020 January 7; in original form 2019 October 10}

\pubyear{2020}

\begin{document}

\label{firstpage}

\pagerange{\pageref{firstpage}--\pageref{lastpage}}

\maketitle


\begin{abstract}
Soon after the recent first ever detection of gravitational waves from merging black holes it has been suggested that their origin is primordial. Appealingly, a sufficient number of primordial black holes (PBHs) could also partially or entirely constitute the dark matter (DM) in the Universe. However, recent studies on PBHs in ultra-faint dwarf galaxies (UFDGs) suggest that they would dynamically heat up the stellar component due to two-body relaxation processes. From the comparison with the observed stellar velocity dispersions and the stellar half-light radii it was claimed that only PBHs with masses $\lesssim10\msun$ can significantly contribute to the DM. In this work, we improve the latter constraints by considering the largest observational sample of UFDGs and by allowing the PBH masses to follow an extended (log-normal) distribution. By means of collisional Fokker--Planck simulations, we explore a wide parameter space of UFDGs containing PBHs. The analysis of the half-light radii and velocity dispersions resulting from the simulations leads to three general findings that exclude PBHs with masses $\sim\mathcal{O}(1$--$100)\msun$ from constituting \textit{all} of the DM: (i) we identify a critical sub-sample of UFDGs that only allows for $\sim\mathcal{O}(1)\msun$ PBH masses; (ii) for any PBH mass, there is an UFDG in our sample that disfavours it; (iii) \JS{the spatial extensions of a majority of simulated UFDGs containing PBHs are too large to match the observed.}
\end{abstract}



\section{Introduction}\label{sec:introduction}

Compelling evidence for the existence of dark matter (DM) has been found on almost all astrophysical scales; still, its very nature  remains elusive. Solving the DM puzzle has therefore become one of the greatest challenges in present-day astrophysics. Recently, the first direct detections of gravitational waves released in black hole binary mergers \citep{2018arXiv181112907T} revived interest in the conjecture that primordial black holes (PBHs) may constitute partly or entirely the DM in the Universe \citep[e.g.][]{2018PDU....22..137C,2016PhRvL.116t1301B,2016PhRvL.117f1101S,2016PhRvD..94h3504C,2017JCAP...09..037R,Garc_a_Bellido_2017,2017PDU....15..142C}. That being the case, numerous PBHs would have formed out of sufficiently large overdensities in the primordial matter power spectrum; such overdensities would eventually have collapsed during the radiation-dominated epoch, producing a population of PBHs that survived to the present \citep{2017JHEAp..13...22R}. The PBH masses at time of formation are expected to be roughly the enclosed mass within the horizon at that time \citep{2018CQGra..35f3001S}. For this reason, depending on the precise formation time, PBHs can span a wide range of masses. Due to \citet{1974Natur.248...30H} radiation, PBHs with masses $M\lsim10^{-19}~\msun{}$ should have already evaporated by now \citep{PhysRevD.13.198}, whereas heavier PBHs could in principle have survived until today, contributing to the present-day DM. Amongst those, PBHs with masses $\mathcal{O}(1$--$100)\msun{}$ are of particular interest as they can account for the astrophysical origin of  BH binaries detected through gravitational waves.

For these PBHs, several observations constrain the possibility that they constitute a significant fraction of the DM. Firstly, if a significant fraction of the Galactic DM halo consists of PBHs, the light coming from background sources would occasionally exhibit a microlensing pattern \citep{1986ApJ...304....1P}. That is, PBHs that intersect the optical axis between observer and source deflect its light rays, resulting in a characteristic magnification of the apparent luminosity over a finite period of time. Several missions aimed at looking for such microlensing events on nearby stars and distant quasars have been carried out in the last decades \citep[e.g.][]{2019NatAs.tmp..238N,2007A&A...469..387T,2001ApJ...550L.169A,2017ApJ...836L..18M}. In total, a few dozen microlensing events have been observed that last up to several months and whose rates put stringent upper limits on PBHs with masses up to $\mathcal{O}(10)\msun{}$ as a dominant DM component \citep{Carr_et_al_2017}. Secondly, a PBH would accrete baryons getting sufficiently close to its horizon; if this happens in the early Universe, the resulting radiation would leave signatures on the power spectrum of the cosmic microwave background (CMB) and on its polarisation anisotropies \citep{2017PhRvD..95d3534A,2018CQGra..35f3001S}. From this, conservative estimates suggest PBHs with masses roughly above $\mathcal{O}(100)\msun{}$ to be excluded from contributing significantly to the DM \citep{Carr_et_al_2017}; however, one should keep in mind that such constraints intrinsically suffer from uncertainties in the modelling of the accretion physics.

In this paper, we investigate whether observations of ultra-faint dwarf galaxies (UFDGs) are compatible with the $\mathcal{O}(1$--$100)$~M$_{\odot}$ PBH scenario. Owing to their extraordinarily high ratio between dynamical mass and stellar mass, UFDGs represent an excellent candidate to test DM models \citep{2019arXiv190105465S}. 

Previous studies of PBHs in UFDGs typically investigated the dynamical imprint of PBHs on the stellar population. \citet{2016ApJ...824L..31B} showed that a recently discovered star cluster in Eridanus~II as well as the entire stellar populations of dwarf galaxies are vulnerable to dynamical heating by PBHs; based on the survival of the star cluster and that of a selected sub-sample of UFDGs, and assuming a monochromatic PBH mass function (i.e. a mass function in which all PBHs have the same mass), he put stringent upper bounds on PBHs with masses respectively $\gsim 5$ and $\gsim 10\msun{}$ as constituting the bulk of DM. \citet{2016PhRvD..94f3530G} generalised these results to an extended (log-normal) PBH mass function by studying the consequences on the Eridanus~II star cluster and on the UFDGs' stellar populations. From this, she excluded broad PBH mass functions with a significant amount of $\mathcal{O}(10)\msun{}$ PBHs. In a different study, \citet{2017PhRvL.119d1102K} investigated effects of mass segregation between PBHs and stars in Segue I. They specifically predicted a depletion of stars in the UFDG's centre as well as a ring structure in the projected stellar surface density profile. By comparing these predictions with the observation of Segue I, they excluded PBHs of single mass $\mathcal{O}(10)\msun{}$ from constituting a significant DM fraction.

Finally, based on a similar sub-sample of UFDGs used by \citet{2016ApJ...824L..31B}, \citet{Zhu_et_al_2018} subsequently aimed to reconstruct the UFDGs' stellar observables via a large number of galaxy simulations with varying the (monochromatic) PBH mass. From the (dis)agreement of observations with the related simulations, they concluded that only PBHs with masses 2--14$\msun{}$ can significantly contribute to the DM. Their simulations also appeared in conflict with the predictions from \citet{2017PhRvL.119d1102K}, as they did not exhibit any ring structure in the stellar density profile.

In summary, dynamical constraints from UFDGs severely challenge the possibility that PBHs would constitute a significant fraction of the DM. Nevertheless, there is room left for improvement of the previous results that could strengthen such dynamical constraints. For instance, previous studies are typically restricted to a limited sample of UFDGs. In fact, only the very faint end of UFDGs has been considered which allows PBH masses around $\mathcal{O}(1)\msun$. In contrast, the growing number of detections invites us to take a larger sample into account. A priori, one would expect three possible outcomes for this more complete investigation. Either the added UFDGs allow the same PBH masses, they allow a different mass window, or they exclude PBHs of any mass from constituting a significant part of the DM. Evidently, the latter two results would severely challenge the PBH scenario. On the other hand, with the exception of \citet{2016PhRvD..94f3530G}, previous studies just considered monochromatic PBH mass functions. However, as pointed out  for instance by \citet{Carr_et_al_2017} and \citet{PhysRevD.92.023524}, this assumption is neither realistic with respect to most of the proposed PBH formation mechanisms nor does it usually lead to the same constraints. Hence, the aim of this work is to overcome this insufficiency by providing a general constraint on the existence of PBHs in UFDGs. For this purpose, we  follow the approach of \citet{Zhu_et_al_2018} but we consider a most extensive sample of UFDGs and an extended PBH mass function. Specifically, we investigate whether PBHs can constitute \textit{all} of the DM (hereafter, PBH-DM).

This paper is organised as follows: Section~\ref{sec:methods} is dedicated to the description of the methods we use. In Section~\ref{sec:results}, we present the results that are discussed in Section~\ref{sec:discussion}.


\section{Methods}\label{sec:methods}

The primary objective of this work is to investigate whether the stellar properties of observed UFDGs are compatible with DM entirely constituted by PBHs. Our analysis is motivated by the fact that the relaxation time can become comparable to the Hubble time in this scenario. Consequently, PBH-DM in UFDGs has to be treated as a collisional fluid potentially affecting the shape of the UFDGs. Such collisional evolution will be reflected in the observable stellar populations.

For this purpose, we perform large sets of $2\times10^5$ \textit{collisional} simulations for each UFDG in our sample (described in Section~\ref{subsec:observation}) and we compare the resulting observational properties with the observed ones. Each simulation models the interplay between stars and PBH-DM altogether specified by a set of input parameters. Furthermore, as opposed to previous work, we do not restrict our analysis to a monochromatic PBH mass function, but also investigate the consequences of an extended (log-normal) PBH mass function (see Section~\ref{subsec:mass-functions}) by explicitly including its model parameters as input parameters. From comparison between simulations and observations, we are thus able to infer the probabilities for each input parameter set to reproduce the stellar properties, i.e. we evaluate if a given set of parameters is reproduced in the associated UFDG or not. Finally, the combined analysis for all UFDGs allows us to impose a generic constraint on PBH-DM.

More specifically, given an UFDG, let $D=\{d_j\}$ and $\Theta=\{\theta_i\}$ be the set of observed stellar properties and the set of input parameters for one simulation, respectively. Then we draw $2\times10^5$ samples from the posterior distribution,

\begin{equation}\label{eq:posterior}
    p(\Theta\mid D)\propto p(\Theta)p(D\mid\Theta),
\end{equation}

\JS{where $p(\Theta)$ and $p(D\mid\Theta)$ are the prior distribution and the likelihood function, respectively}, by the use of {\fontfamily{qcr}\selectfont emcee} \citep[][]{2013PASP..125..306F}, an implementation of the affine invariant ensemble sampler for Markov chain Monte Carlo (MCMC) methods proposed by \citet{2010CAMCS...5...65G}. Whenever the ensemble sampler calls the likelihood function $p(D\mid\Theta)$, a whole UFDG is initialised by means of $\Theta$, subsequently simulated over an extended period of time $t$, and finally compared against $D$.

To perform the simulations of the UFDGs, we use {\fontfamily{qcr}\selectfont PhaseFlow} \citep[][]{2017ApJ...848...10V}, which is part of the publicly available software library {\fontfamily{qcr}\selectfont Agama} \citep[][]{2019MNRAS.482.1525V}. {\fontfamily{qcr}\selectfont PhaseFlow} dynamically evolves a given spherically symmetric system consisting of one or more {\it collisional} components (e.g. stars, stellar-mass BHs, supermassive BHs, etc.) by simultaneously solving the coupled set of Poisson and Fokker--Planck equations, incorporating the effect of two-body relaxation processes. Its low computational cost allows us to perform the large sets of simulations required for the exploration of the input parameters using MCMC. Additionally, {\fontfamily{qcr}\selectfont PhaseFlow} is well tested in the context of nuclear star clusters \citep{Generozov2018,Emami2019,Emami2019b}, Bahcall--Wolf cusps \citep[][]{Bahcall1976,2017ApJ...848...10V}, and monochromatic PBH mass functions \citep[][]{Zhu_et_al_2018}.

In the following, we explicate each aspect relevant for our approach, including the observational sample of UFDGs (Section~\ref{subsec:observation}), the selection of the PBH mass functions (Section~\ref{subsec:mass-functions}), the models of UFDGs (Section~\ref{subsec:simulation}), and the choice of the prior distribution $p(\Theta)$ as it appears in Eq.~\eqref{eq:posterior} (Section~\ref{subsec:prior}).

\subsection{Sample of observed UFDGs}\label{subsec:observation}

Following the definition of \cite{2019arXiv190105465S}, UFDGs are dwarf galaxies with V-band luminosities $L_\text{V} \leq 10^5$~L$_{\odot}$, corresponding to absolute magnitudes $M_\text{V}\geq-7.7$. Considering their stellar masses, surface brightnesses, sizes, dynamical masses, and metallicities, the hitherto observed UFDGs do not appear as fundamentally distinct from other dwarf galaxies \citep{2019arXiv190105465S}. Unlike in previous studies on PBH-DM in UFDGs, we therefore do not restrict ourselves to a selected, seemingly distinct sub-sample of UFDGs \citep[as done in, e.g.][]{2016ApJ...824L..31B,Zhu_et_al_2018}, but instead consider an extensive sample of $27$ UFDGs whose projected two-dimensional stellar half-light radii ($r_{\rm h,\star}$) and \JS{line-of-sight} stellar velocity dispersions ($\sigma_\star$) have been observationally constrained. We list them in Table~\ref{tab:sample}. We assume that each star in the simulation shares the same mass and the same mass-to-light ratio (cf. Section~\ref{subsec:simulation}). Consequently, the (projected two-dimensional) half-light radius is equivalent to the (projected two-dimensional) half-mass radius of the stars. In order to resolve both the stellar kinematics and their structural properties, we base the comparison between simulations and observations on the stellar velocity dispersion and the deprojected stellar half-mass radius $R_{\rm h,\star}=(4/3)r_{\rm h,\star}$ \citep{2010MNRAS.406.1220W}, i.e. $D=\{\sigma_\star,R_{\rm h,\star}\}$. Thus, if a simulation of a given UFDG observed with ${(\sigma_{\rm \star,D})}^{+\sigma_1}_{-\sigma_2}$ and ${(R_{\rm h,\star,D})}^{+\sigma_3}_{-\sigma_4}$ (cf. Table \ref{tab:sample}) outputs the stellar velocity dispersion $\sigma_{\rm \star,S}$ and the deprojected half-mass radius $R_{\rm h,\star,S}$, its likelihood function as it appears in Eq.~\eqref{eq:posterior} is evaluated to

\begin{equation}\label{eq:split-normal}
p(D\mid\Theta)\propto\mathcal{SN}(\sigma_{\rm \star,S};\sigma_{\rm \star,D},\sigma_1,\sigma_2)\mathcal{SN}(R_{\rm h,\star,S};R_{\rm h,\star,D},\sigma_3,\sigma_4),      \end{equation}

where $\mathcal{SN}$ denotes the split-normal distribution. The $\sigma_i$ ($i = 1$--4) are the respective measurement uncertainties. This approach accounts for the asymmetry in some of the reported measurements. For those UFDGs where only an upper limit for $\sigma_{\star}$ is given, $\mathcal{SN}$ in Eq.~\eqref{eq:split-normal} is replaced by a half-normal distribution whose parameter is chosen such that its cumulative distribution function evaluated at the given upper limit matches the respective confidence level.

Lastly, when analysed, the UFDGs' stellar populations turn out to be fairly old ($t \approx 9$--$14$~Gyr). Therefore, we run each simulation over $t = 12$~Gyr.

\subsection{Extended PBH mass functions}\label{subsec:mass-functions}

The two key improvements of this paper compared to \citet{Zhu_et_al_2018} are the generalisations to a severely enlarged sample of UFDGs, presented in Section \ref{subsec:observation}, and to PBHs covering an extended mass range. Usually, the latter are described by means of a mass function $\psi(M)$, which determines their mass density $\rho_\text{PBH}$ and number density $n_\text{PBH}$:

\begin{align}
    \frac{\text{d}\rho_\text{PBH}}{\text{d}M}&=\rho_\text{DM}\psi(M),\label{eq:mass-density}\\
    \frac{\text{d}n_\text{PBH}}{\text{d}M}&=\rho_\text{DM}\frac{\psi(M)}{M}.\label{eq:number-density}
\end{align}

Here, $\rho_\text{DM}$ denotes the total DM mass density. So defined, the mass function is normalised to the total fraction $f_\text{tot} \equiv \rho_\text{PBH}/\rho_\text{DM}$:

\begin{equation}\label{eq:normalisation}
    f_\text{tot} = \int_0^\infty{\psi(M)\,dM}.
\end{equation}

Evidently, $0 \leq f_\text{tot} \leq 1$ must hold. In this paper, we investigate whether PBHs can constitute \textit{all} of the DM, i.e. whether $f_\text{tot}=1$ or not.\footnote{The method we use in this paper is not suited to investigate DM models that \textit{partly} consist of PBHs ($f_\text{tot}<1$), if the complementary components are collisionless, as e.g. weakly interacting massive particles [cf. \citet{2019arXiv190108528A}]; this is because {\fontfamily{qcr}\selectfont PhaseFlow} cannot handle collisionless components.} Henceforth, we set $f_\text{tot}=1$ and $\rho_\text{PBH}$ and $\rho_\text{DM}$ can be used equivalently. By mean of Eqs~\eqref{eq:mass-density}, \eqref{eq:number-density}, and \eqref{eq:normalisation}, we introduce the mass function $\psi(M)$ such that it agrees with the notation of \citet{Carr_et_al_2017}, who comprehensively generalised previously existing constraints on monochromatic mass functions to extended. In the following, we list the functional forms of both the monochromatic mass function and the log-normal mass function, which we investigate as a commonly discussed representative for generic extended mass functions.

\begin{landscape}
\begin{table}
\vspace{30pt}
\caption{List of observed UFDGs with their absolute V-band magnitudes, V-band luminosities, projected two-dimensional stellar half-light radii, and stellar velocity dispersions [cf. \citet{2019arXiv190105465S}]. From top to bottom, the UFDGs are sorted from the faintest to the most luminous. The names of the five UFDGs used by \citet{2016ApJ...824L..31B} and \citet{Zhu_et_al_2018} are highlighted in bold.}
\begin{threeparttable}
\label{tab:sample}
\begingroup
\renewcommand*{\arraystretch}{1.2}
\begin{tabular}{lcccccc}
\hline
\hline
UFDG Name & $M_\text{V}$ & $L_\text{V}$ & $r_{\rm h,\star}$ & $\sigma_\star$ & References for quantities\\
& mag & ${\rm L}_{\sun}$ & pc & km~s$^{-1}$ & $M_\text{V}$ (1), $L_\text{V}$ (2), $r_{\rm h,\star}$ (3), and $\sigma_\star$ (4)\\
\hline
Draco II & $-0.8^{+0.4}_{-1.0}$ & $1.8^{+1.2}_{-0.7}\times10^2$ & $19.0^{+4.5}_{-2.6}$ & $<5.9$ ($95\%$ C.L.) $\tnotex{tn:statistics}$ & All: \citet{2018MNRAS.480.2609L} \\ 

\textbf{Segue I} & $-1.30\pm 0.73$ & $2.8^{+2.7}_{-1.4}\times10^2$ & $24.2\pm2.8$ & $3.7^{+1.4}_{-1.1}$ & 1--3: \citet{2018ApJ...860...66M}, 4: \citet{2011ApJ...733...46S}\\

Tucana III & $-1.3\pm0.2$ & $2.8^{+0.6}_{-0.5}\times10^2$ & $34\pm8$ & $<1.2$ ($90\%$ C.L.) $\tnotex{tn:statistics}$ & 1--3: \citet{2018ApJ...863...25M}, 4: \citet{2017ApJ...838...11S}\\

Triangulum II & $-1.8\pm0.5$ & $4.5^{+2.6}_{-1.7}\times10^2$ & $17.4\pm4.3$ & $<3.4$ ($90\%$ C.L.) $\tnotex{tn:statistics}$ & 1,2: \citet{2015ApJ...802L..18L}, 3: \citet{2018ApJ...860...66M}, 4: \citet{2017ApJ...838...83K}\\

\textbf{Segue II} & $-1.86\pm0.88$ & $4.7^{+6.9}_{-1.6}\times10^2$ & $38.3\pm2.8$ & $<2.6$ ($95\%$ C.L.) $\tnotex{tn:statistics}$ & 1--3: \citet{2018ApJ...860...66M}, 4: \citet{2013ApJ...770...16K}\\

Carina III & $−2.4\pm0.2$ & $7.8^{+1.6}_{-1.3}\times10^2$ & $30\pm9$ & $5.6^{+4.3}_{-2.1}$ $\tnotex{tn:statistics}$ & 1--3: \citet{2018MNRAS.475.5085T}, 4: \citet{2018ApJ...857..145L}\\

\textbf{Willman I} & $-2.53\pm0.74$ & $8.8^{+8.6}_{-4.3}\times10^2$ & $27.7\pm2.4$ & $4.0\pm0.8$ & 1--3: \citet{2018ApJ...860...66M}, 4: \citet{2011AJ....142..128W}\\

Bo\"{o}tes II & $-2.94\pm0.74$ & $1.3^{+1.3}_{-0.6}\times10^3$ & $ 38.7\pm5.1$ & $10.5\pm7.4$ & 1--3: \citet{2018ApJ...860...66M}, 4: \citet{2009ApJ...690..453K}\\

Grus I & $-3.47\pm0.59$ & $2.1^{+1.5}_{-0.9}\times10^3$ & $28.3\pm23.0$ & $2.9^{+6.9}_{-2.1}$ & 1--3: \citet{2018ApJ...860...66M}, 4: \citet{2016ApJ...819...53W}\\

\textbf{Horologium I} & $-3.55\pm0.56$ & $2.2^{+1.5}_{-0.9}\times10^3$ & $36.5\pm7.1$ & $4.9^{+2.8}_{-0.9}$ & 1--3: \citet{2018ApJ...860...66M}, 4: \citet{2015ApJ...811...62K}\\

\textbf{Reticulum II} & $-3.88\pm0.38$ & $3.0^{+1.3}_{-0.9}\times10^3$ & $48.2\pm1.7$ & $3.3\pm0.7$ & 1--3: \citet{2018ApJ...860...66M}, 4: \citet{2015ApJ...808...95S}\\

Tucana II & $-3.9\pm0.2$ & $3.1^{+0.6}_{-0.5}\times10^3$ & $120\pm30$ & $8.6^{+4.4}_{-2.7}$ & 1--3: \citet{2015ApJ...807...50B}, 4: \citet{2016ApJ...819...53W}\\

Pegasus III & $-4.1\pm0.5$ & $3.7^{+2.2}_{-1.4}\times10^3$ & $78^{+30}_{-24}$ & $5.4^{+3.0}_{-2.5}$ & 1--3: \citet{2015ApJ...804L..44K}, 4: \citet{2016ApJ...833...16K}\\

Pisces II & $-4.22\pm0.38$ & $4.2^{+1.7}_{-1.2}\times10^3$ & $59.3\pm8.5$ & $5.4^{+3.6}_{-2.4}$ & 1--3: \citet{2018ApJ...860...66M}, 4: \citet{2015ApJ...810...56K}\\

Ursa Major II & $-4.25\pm0.26$ & $4.3^{+1.2}_{-0.9}\times10^3$ & $128\pm5$ & $5.6\pm1.4$ & 1--3: \citet{2018ApJ...860...66M}, 4: \citet{2019arXiv190105465S}\\

Aquarius II & $-4.36\pm0.14$ & $4.7^{+0.7}_{-0.6}\times10^3$ & $159\pm24$ & $5.4^{+3.4}_{-0.9}$ & 1--4: \citet{2016MNRAS.463..712T}\\

Coma Berenices & $-4.38\pm0.25$ & $4.8^{+1.3}_{-1.0}\times10^3$ & $72.1\pm3.8$ & $4.6\pm0.8$ & 1--3: \citet{2018ApJ...860...66M}, 4: \citet{2007ApJ...670..313S}\\

Leo V & $-4.40\pm0.36$ & $4.9^{+1.9}_{-1.4}\times10^3$ & $51.8\pm16.6$ & $2.3^{+3.2}_{-1.6}$ $\tnotex{tn:statistics}$ & 1--3: \citet{2018ApJ...860...66M}, 4: \citet{2017MNRAS.467..573C}\\

Carina II & $-4.5\pm0.1$ & $5.4^{+0.5}_{-0.5}\times10^3$ & $91\pm8$ & $3.4^{+1.2}_{-0.8}$ & 1--3: \citet{2018MNRAS.475.5085T}, 4: \citet{2018ApJ...857..145L}\\

Hydra II & $-4.60\pm0.37$ & $5.9^{+2.4}_{-1.7}\times10^3$ & $ 59.2\pm10.9$ & $<3.6$ ($90\%$ C.L.) $\tnotex{tn:statistics}$ & 1--3: \citet{2018ApJ...860...66M}, 4: \citet{2015ApJ...810...56K}\\

Hydrus I & $-4.71\pm0.08$ & $6.5^{+0.5}_{-0.5}\times10^3$ & $53\pm4$ & $2.7\pm0.5$ & All: \citet{2018MNRAS.479.5343K}\\

Leo IV & $-4.99\pm0.26$ & $8.5^{+2.3}_{-1.8}\times10^3$ & $114\pm12$ & $3.3\pm1.7$ & 1--3: \citet{2018ApJ...860...66M}, 4: \citet{2007ApJ...670..313S}\\

Ursa Major I & $-5.12\pm0.38$ & $9.6^{+4.0}_{-2.8}\times10^3$ & \JS{$293.4^{+34.4}_{-32.6}$} & $7.0\pm1.0$ & 1,2: \citet{2018ApJ...860...66M}, \JS{3: \citet{Oka}}, 4: \citet{2019arXiv190105465S}\\

Canes Venatici II & $-5.17\pm0.32$ & $1.0^{+0.3}_{-0.3}\times10^4$ & $70.7\pm11.2$ & $4.6\pm1.0$ & 1--3: \citet{2018ApJ...860...66M}, 4: \citet{2007ApJ...670..313S}\\

Hercules & $-5.83\pm0.17$ & $1.8^{+0.3}_{-0.3}\times10^4$ & $216\pm17$ & $5.1\pm0.9$ & 1--3: \citet{2018ApJ...860...66M}, 4: \citet{2007ApJ...670..313S}\\

Bo\"{o}tes I & $-6.02\pm0.25$ & $2.2^{+0.6}_{-0.5}\times10^4$ & $191\pm5$ & $2.4^{+0.9}_{-0.5}$ $\tnotex{tn:Bootes I}$ &  1--3: \citet{2018ApJ...860...66M}, 4: \citet{2011ApJ...736..146K}\\

Eridanus II & $-7.1\pm0.3$ & $5.9^{+1.9}_{-1.4}\times10^4$ & $277\pm14$ & $6.9^{+1.2}_{-0.9}$ & \JS{1--3}: \citet{2016ApJ...824L..14C}, 4: \citet{2017ApJ...838....8L}\\

\hline
\hline
\end{tabular}
\endgroup
\begin{tablenotes}
\item [$\spadesuit$] \label{tn:statistics} The value for $\sigma_\star$ suffers from poor statistics based on only few stars (see respective reference). Therefore, caution is advised when interpreting the inferred results.
\item [$\clubsuit$] \label{tn:Bootes I} \citet{2011ApJ...736..146K} discovered a major, \say{cold} stellar population with $\sigma_\star=2.4^{+0.9}_{-0.5}\,$km~s$^{-1}$, and a second, minor, \say{hot} population with $\sigma_\star\sim9\,$km~s$^{-1}$. For our analysis, we adopt the former value.
\end{tablenotes}
\end{threeparttable}
\end{table}
\end{landscape}

\begin{itemize}[leftmargin=.2in]

    \item A monochromatic mass function:
    
    \begin{equation}\label{eq:monochromatic}
        \psi\left(M;M_{\rm c}\right)=\delta\left(M-M_{\rm c}\right),
    \end{equation}
    
    where $\delta$ denotes the Dirac delta function. By definition, this mass function only depends on the PBH mass $M$. Typical mass functions emerging from a critical collapse of density fluctuations with a $\delta$-function power spectrum are relatively narrow and can therefore also be well approximated with a monochromatic mass function \citep{Carr_et_al_2017}. When performing the simulations, we adopt the PBH mass as a free input parameter $M_{\rm c}\in\Theta$.\\

    \item A log-normal mass function:
    
    \begin{equation}\label{eq:log-normal-0}
        \psi(M;\mu,\sigma)=\frac{1}{M\sqrt{2\pi\sigma^2}}\exp{\left(- \frac{\ln^2(M/\mu)}{2\sigma^2}\right)},
    \end{equation}
    
    \noindent where $\mu > 0$ and $\sigma > 0$. The mean PBH mass in the assumption of a log-normal mass function is $\bar{M}=\mu\exp\left(-\sigma^2/2\right)$. Such mass function is a good approximation for a large class of PBH formation scenarios, e.g. axion-curvaton, running-mass, and single field double inflation \citep{2016PhRvD..94f3530G,2017JCAP...09..020K}. The  $(\mu, \sigma)$ pairs allowed in the window relevant for the operational gravitational wave detectors [$\mathcal{O}(1$--$100)\,{\rm M}_{\odot}$; \citealt{2018arXiv181112907T}] are mainly constrained by (i) the results from searches for microlensing events on stars in our Galactic neighbourhood and (ii) the effect PBH gas accretion would have on the CMB temperature and ionization history \citep{2018CQGra..35f3001S,Carr_et_al_2017}. Together, the allowed points roughly constitute a sub-plane described as \citep[see figure~3 in][]{Carr_et_al_2017}:
    
    \begin{equation}\label{eq:mu-sigma-plane}
        (\mu,\sigma)\in[25\,{\rm M}_\odot,100\,{\rm M}_\odot]\times[0.0,1.0].
    \end{equation}
    
    It is worth mentioning that these points are ruled out completely if one further takes the following constraints into account \citep{Carr_et_al_2017}: (iii) the survival of the stellar cluster in Eridanus II and of the entire stellar populations in UFDGs \citep{2016ApJ...824L..31B,2017PhRvL.119d1102K,2016PhRvD..94f3530G,Zhu_et_al_2018} and (iv) the survival of wide binaries in the Milky Way \citep{2014ApJ...790..159M}. These constraints are somewhat less restrictive as they rely on further astrophysical assumptions \citep{Carr_et_al_2017}. For instance, the stellar cluster in Eridanus II could have only recently spiralled down into the centre of the galaxy, where dynamical heating by PBHs becomes effective \citep{2016ApJ...824L..31B}. Additionally, wide binaries are in principle hard to detect, yielding some uncertainty in identifying them and consequently in drawing any conclusion on the PBH abundance \citep{2018CQGra..35f3001S}. Moreover, we emphasise that, most recently, even previous microlensing constraints were called into question as spatial PBH clustering \citep{2018PDU....19..144G} and updated  galactic rotation curves \citep{2015A&A...575A.107H} tend to relax them.
    
    When we implement the log-normal mass function in the simulation, we take both parameters, $\mu$ and $\sigma$, as free input parameters: $\mu,\sigma\in\Theta$.
    
\end{itemize}

\subsection{Initial UFDG configurations}\label{subsec:simulation}

\begin{figure}
\vspace{-5pt}
\centering
\includegraphics[width=0.45\textwidth]{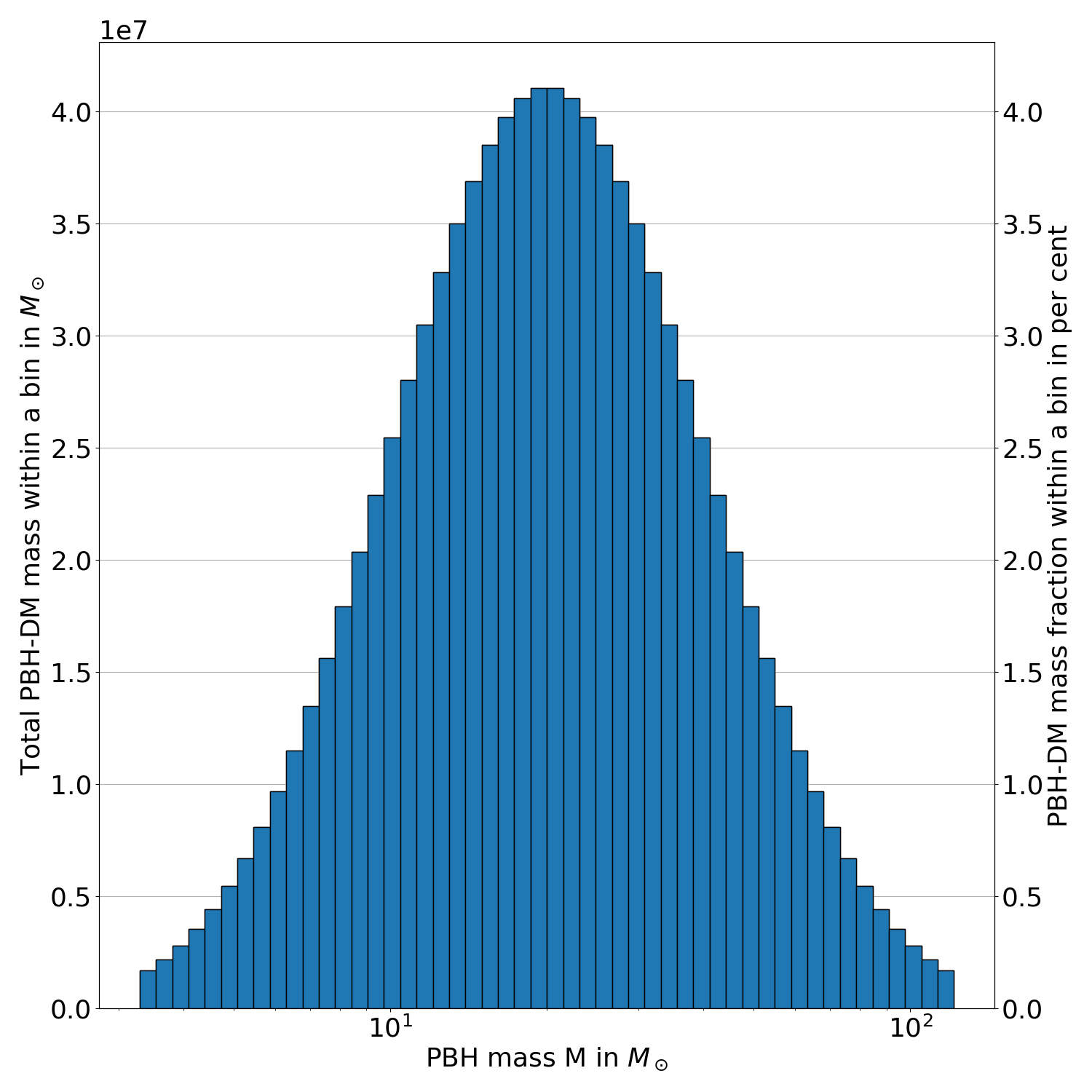}
\caption{Binning scheme for a log-normal PBH mass function with $\mu=20~\msun{}$ and $\sigma=0.7$. The total DM mass is set to $M_\text{DM}=10^9\,{\rm M}_\odot$. The $1$ per cent of the mass in the tails of the distribution is neglected. Thus, the $N=50$ bins are limited by the thresholds  $M_\text{th,0}\approx 3.3\,{\rm M}_\odot$ and $M_\text{th,1}\approx 121.4\,{\rm M}_\odot$ and in total make up 99 per cent of the total DM mass. Each bin corresponds to one PBH-DM component in the simulation with a PBH mass approximated with the average PBH mass within the bin, cf. Eq.~\eqref{eq:average}.}
\vspace{-5pt}
\label{fig:log-normal}
\end{figure}

Each point in the MCMC represents an UFDG that consists of stars and PBHs and whose evolution is simulated over an integration time $t=12\,\text{Gyr}$ with {\fontfamily{qcr}\selectfont PhaseFlow}. The numerical value for $t$ is motivated in Section~\ref{subsec:observation} by the given observational UFDG sample. For any simulation, we assume that the initial radial density distribution of the PBHs is described by a \citet{1993MNRAS.265..250D} sphere:

\begin{equation}\label{eq:dehnen}
    \rho_\text{PBH}(r)=\frac{(3-\gamma)M_\text{DM}}{4\pi R^3_{0,\text{DM}}}\left(\frac{r}{R_{0,\text{DM}}}\right)^{-\gamma}\left(1+\frac{r}{R_{0,\text{DM}}}\right)^{\gamma-4},
\end{equation}

where $M_\text{DM}$, $R_{0,\text{DM}}$, and $\gamma$ are the DM total mass, scale radius, and asymptotic inner slope, respectively. Whilst taking $M_\text{DM}$ and $R_{0,\text{DM}}$ as free input parameters ($M_\text{DM},R_{0,\text{DM}}\in\Theta$), we fix $\gamma=0$ for all simulations, as PBHs have been shown \citep[][]{Zhu_et_al_2018} to be a cusp-core transformer by rapidly converting an initially cuspy ($\gamma=1$) density profile into a cored one ($\gamma=0$). We identified the same behaviour in direct $N$-body simulations of dwarf galaxies with PBH-DM that is initially described by a cuspy density profile. Most recently, this result was also found by \citet{2019arXiv190907395B}. In their simulations, \citet{Zhu_et_al_2018} implement a single DM component consisting of monochromatic PBHs that follow Eq.~\eqref{eq:dehnen}. Here we also investigate the extended mass functions introduced in Section~\ref{subsec:mass-functions} by approximating them with $N=50$ mass bins, each of which corresponds to one component in the simulations. We emphasise that this generalisation is a main improvement presented in this paper since it allows us to constrain physically more realistic PBH mass functions than the monochromatic one. More specifically, given a mass function $\psi(M)$,

\begin{equation}
    F(0;M_\text{th})=\int_0^{M_\text{th}}{\psi(M)\,{\rm d}M}
\end{equation}

is the cumulative DM fraction of PBHs with masses $M\in[0,M_\text{th}]$ [cf. with Eq.~\eqref{eq:normalisation}]. For all mass functions introduced in Section~\ref{subsec:mass-functions}, $F(0;M_\text{th})$ is invertible so that there are uniquely determinable mass thresholds $M_\text{th,0}$ and $M_\text{th,1}$ at which $F(0;M_\text{th,0})=0.5$ per cent and $F(0;M_\text{th,1})=99.5$ per cent, respectively. Once $M_\text{th,0}$ and $M_\text{th,1}$ are calculated, we divide the mass range in $\log_{10}$-space, $[\log_{10} M_\text{th,0},\log_{10} M_\text{th,1}]$, in $N$ bins of same length and assume that all PBHs in one bin have a mass equal to the bin's average, 

\begin{equation}\label{eq:average}
    \bar{M}=\frac{\int_{M_\text{bin,0}}^{M_\text{bin,1}}\psi(M)\,{\rm d}M}{\int_{M_\text{bin,0}}^{M_\text{bin,1}}(\psi(M)/M)\,{\rm d}M},
\end{equation}

where $M_\text{bin,0}$ and $M_\text{bin,1}$ are the bin's upper and lower boundary, respectively. Consequently, the total DM mass contained within a bin is computed as its fractional mass, $M_\text{DM} F(M_\text{bin,0};M_\text{bin,1})$. Loosely speaking, we thus perform the transition from a monochromatic to an extended mass function by replacing the single monochromatic PBH-DM component with $N$ PBH-DM components, each of which is in turn monochromatic but at different masses dictated by the given mass function. In this approach, we artificially neglect $1$ per cent at the tails of the mass function, primarily in order to avoid bins with an average PBH mass greater than the bin's total DM mass, which would otherwise lead to numerical errors in the simulation. In Figure~\ref{fig:log-normal}, this mass binning scheme is visualised for an exemplary mass function.

Concerning the initial distribution of the stars, we set up a \citet{1911MNRAS..71..460P} sphere, described by

\begin{equation}\label{eq:plummer}
      \rho_\star(r)=\frac{3M_\star}{4\pi R^3_{0,\star}}\left( 1+\frac{r^2}{R^2_{0,\star}} \right)^{-5/2},
  \end{equation}
  
where $M_\star$ and $R_{0,\star}$ are the total stellar mass and scale radius, respectively. Here, $R_{0,\star}$ is taken as a free input parameter ($R_{0,\star}\in\Theta$). The total stellar mass $M_\star$ is estimated from the luminosity of each UFDG separately (cf. Table \ref{tab:sample}) by assuming a stellar mass-to-light ratio of 2~M$_{\odot}$/L$_{\odot}$, since the UFDGs' stars are fairly old \citep{2019arXiv190105465S}. We also perform comparison runs with 1 and 3~M$_{\odot}$/L$_{\odot}$. Additionally, the stellar mass distribution is assumed to be monochromatic at $1\msun{}$. Even though assuming a single mass value for all the stars is a simplification, we do not expect it to impact our results as (i) stellar masses cover a less broad mass range compared to the one tested for PBHs and (ii) the low mass of stars, compared to the PBH mass obtained in this study, implies that the overall dynamical evolution (relaxation) of the simulated UFDGs is virtually not influenced by the stellar population.

Finally, the average velocity changes per unit time experienced by the stars and PBHs due to mutual gravitational interactions are typically described by means of diffusion coefficients. These scale with the Coulomb logarithm $\ln\Lambda$, where $\Lambda=b_\text{max}/b_{90}$ is the ratio between the maximum impact parameter, $b_\text{max}$, and the impact parameter, $b_{90}$, for which 90 degree scattering occurs \citep{2008gady.book.....B}. Unfortunately, there is no precise definition for either $b_\text{90}$ or $b_\text{max}$, as they can be somehow position-dependent for a given system. In fact, $\ln\Lambda$ is not straightforwardly defined even for the simplest possible scenario of homogeneously distributed perturbers of the same mass \citep{2013degn.book.....M}. Therefore, we test three typical values for the Coulomb logarithm, i.e. $\ln\Lambda=5$, $10$, and $15$.

\subsection{Prior distributions of the input parameters}\label{subsec:prior}

In total, a set of input parameters $\Theta$ consists of the DM scale radius $R_\text{0,DM}$, the stellar scale radius $R_{0,\star}$, the total DM mass $M_\text{DM}$, and one or two parameters that characterise the assumed PBH mass function, i.e. $\Theta=\{R_{0,\star},R_\text{0,DM},M_\text{DM},M_{\rm c}\}$ and  $\Theta=\{R_{0,\star},R_\text{0,DM},M_\text{DM},\mu,\sigma\}$ for the monochromatic and  log-normal PBH mass function, respectively. Their prior distributions $p(\theta_i)$ are assumed to be independent of each other, so that the joint prior distribution $p(\Theta)$ in Eq.~\eqref{eq:posterior} is just the product of the individual priors:

\begin{equation}\label{eq:product}
    p(\Theta)=\prod_i{p(\theta_i)}.
\end{equation}

In order to reflect the increased variety of our observational sample (Table~\ref{tab:sample}) compared to the one used by \citet[][]{Zhu_et_al_2018}, we uninformatively consider log-uniform distributions over enlarged ranges for the individual priors of $R_\text{0,DM}$ and $R_{0,\star}$:

\begin{align}
    \log_{10}&(R_\text{0,DM}/\text{pc})\sim\mathcal{U}(\log_{10}100;\log_{10}20000),\label{eq:parameter-DM}\\
    \log_{10}&(R_{0,\star}/\text{pc})\sim\mathcal{U}(\log_{10}10;\log_{10}300).
\end{align}

Concerning the total DM mass $M_\text{DM}$, we adopt the log-normal prior distribution from \citet[][]{Zhu_et_al_2018}, as it covers the most recent DM halo mass estimates:

\begin{align} \label{eq:halo_prior}
    \log_{10}&(M_\text{DM}/{\rm M}_\odot)\sim\mathcal{N}(9.0;0.5).
\end{align}

The remaining input parameters depend on the specific choice of mass function (cf. Section \ref{subsec:mass-functions}), and we assume log-uniformity for both the monochromatic mass function,
    
    \begin{equation}
        \log_{10}(M_{\rm c}/{\rm M}_\odot)\sim\mathcal{U}(\log_{10}1;\log_{10}400),
    \end{equation}
    
and the log-normal mass function,
    
    \begin{align}
        &\log_{10}(\mu/{\rm M}_\odot)\sim\mathcal{U}(\log_{10}1;\log_{10}400),\\
        &\log_{10}(\sigma)\sim\mathcal{U}(\log_{10}0.1;\log_{10}1.5).
    \end{align}
    
Thus, Eqs~\eqref{eq:split-normal} and \eqref{eq:product} can be used to determine the posterior distribution $p(\Theta\mid D)$ via Eq. \eqref{eq:posterior}, that measures the likelihood of a simulated UFDG to meet the observations.


\section{Results}\label{sec:results}

\subsection{Monochromatic mass function}\label{sec:mono-results}

\begin{figure*}
\vspace{-5pt}
\centering
\includegraphics[width=1.0\textwidth]{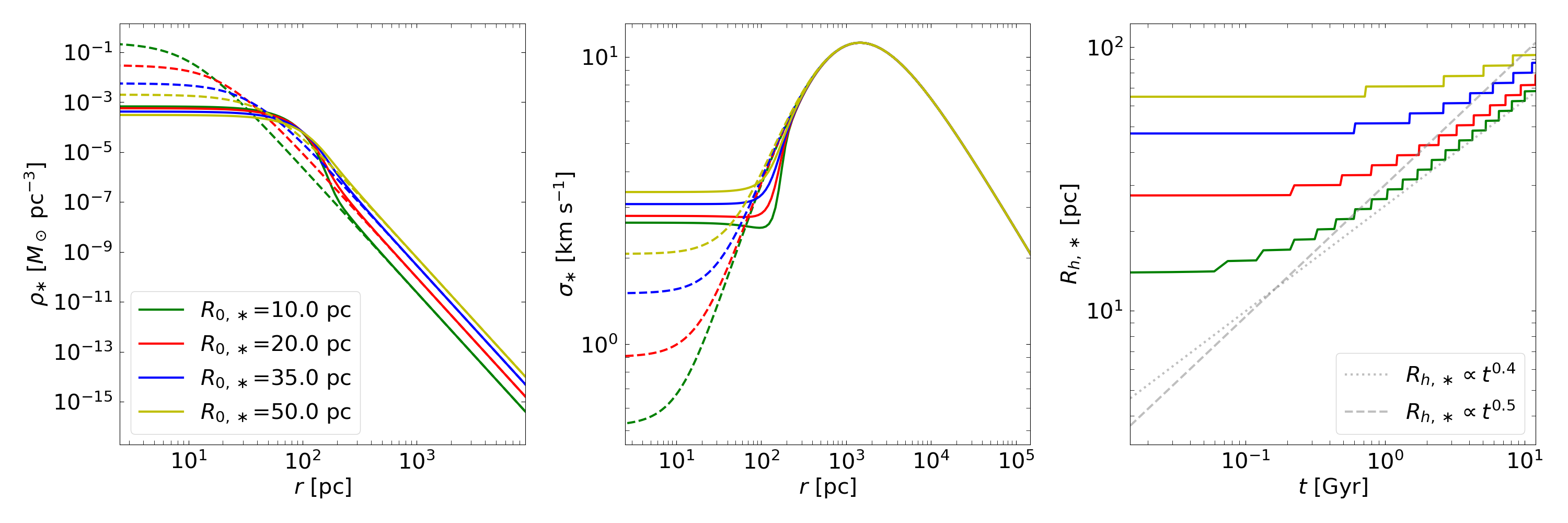}
\caption{Stellar mass density profile (left-hand panel), stellar velocity dispersion profile (central panel), and temporal evolution of the deprojected half-mass radius $R_{\rm h,\star}$ (right-hand panel) for four runs with a monochromatic PBH mass function ($M_{\rm c}=30\msun{}$), a total DM mass of $M_\text{DM}=10^9\msun{}$, a total stellar mass of $M_\star=10^3\msun$, and a DM scale radius $R_\text{DM,0}=10^3$~pc, which differ only in the initial stellar scale radius: $R_{0,\star} = 10$ (green), $20$ (red), $35$ (blue), and $50\,\text{pc}$ (yellow). The left-hand and central panels show the stellar profiles at the beginning ($t = 0$; dashed lines) and at the end of the simulation ($t = 12$~Gyr; solid lines). The grey dotted and dashed lines in the right-hand panel show the reference growth rates $R_{\rm h,\star}\propto t^{0.4}$ and $R_{\rm h,\star}\propto t^{0.5}$, respectively.
}
\vspace{-5pt}
\label{fig:mono-halfmass-evo}
\end{figure*}

In this section, we describe the dynamical effects of monochromatic PBHs on the stellar population of the UFDGs. For this purpose, we perform four fiducial simulations varying the stellar scale radius $R_{0,\star} = \{10, 20, 35, 50 \}$~pc, while fixing the mass of each PBH to $M_{\rm c}=30 \msun{}$, the total stellar mass to  $M_\star=10^3\msun{}$, the total DM mass to $M_\text{DM}=10^9\msun{}$, and the DM scale radius to $R_\text{DM,0}=10^3\,\text{pc}$. We fix the Coulomb logarithm $\ln\Lambda=15$ for all of these simulations.

The resulting stellar mass density, $\rho_\star(r)$, and velocity dispersion, $\sigma_\star(r)$, profiles at the beginning and end of the simulation, as well as the temporal evolution of the deprojected half-mass radius $R_{\rm h,\star}(t)$ are shown in Figure~\ref{fig:mono-halfmass-evo}. The panels show that the central mass density of stars drops with time, while the central stellar velocity dispersion increases. This behaviour can be explained in terms of the initial temperature inversion of stars, i.e. the fact that their velocity dispersion exhibits a pronounced peak at some radius at the beginning of the integration (in our specific case, at $\sim$1~kpc). In other words, one can think of the stars at the very centre as a dynamically \say{cool} sub-system \citep[][]{Binney_1980}. As time progresses, gravitational encounters due to PBHs and other stars increase the central stellar velocity dispersion. As a result, the central stars get dynamically heated up and  expand to larger radii. This phenomenon is well studied for the case of  compact stellar systems lacking a central massive black hole, e.g. nuclear star clusters \citep{1996NewA....1..255Q}. The expansion of the stellar population results in an enhancement of the stellar half-mass radius (right-hand panel in Figure~\ref{fig:mono-halfmass-evo}). This phenomenon was analytically described by \citet{2016ApJ...824L..31B}: the stellar system slowly expands until it gets completely DM-dominated at all radii, while the half-mass radius grows with time as $R_{\rm h,\star}\propto t^{0.5}$. Our simulations yield a slightly less efficient heating rate for initial stellar scale radii $R_{0,\star}$ between 10 and 50~pc, as previously found by \citet{Zhu_et_al_2018} for $R_{0,\star}$ up to $15\,\text{pc}$. The reason for this could be that two dynamical effects of the PBHs on the stellar population were implicitly neglected in the original derivation of the analytic expression by \citet{2016ApJ...824L..31B}: on the one hand, dynamical friction of the stars due to the PBHs cools the stellar population; on the other hand, the increase of the stellar velocity dispersion should gradually lower the heating efficiency by PBHs. If both effects are taken into account, heating would be, in principle, less efficient and the stellar population would expand at a lower pace than anticipated. In order to investigate whether these effects are significant or not, we calculate the complete heating-to-cooling ratio in the inner $10^2\,\text{pc}$ at the initial time, $t=0$, and final time, $t=12\,\text{Gyr}$. That is, we evaluate the fraction,

\begin{align}
\frac{\text{heating}}{\text{cooling}}&=-\frac{D[(\Delta v_\parallel)^2]+D[(\Delta \bm{v}_\bot)^2]}{2vD[\Delta v_\parallel]}\\
&=\frac{M_{\rm c}}{m_\star+M_{\rm c}}\frac{\text{erf}(X)}{2X^2G(X)},
\end{align}

for all of the four simulations as a function of radius, where $D[(\Delta v_\parallel)^2$, $D[(\Delta \bm{v}_\bot)^2]$, and $D[\Delta v_\parallel]$ (dynamical friction) are the diffusion coefficients for the stars that are assumed to follow a Maxwellian velocity distribution, $X\equiv v/(\sqrt{2}\sigma_\text{PBH})$ is the ratio between the typical stellar velocity $v$ and the velocity dispersion $\sigma_\text{PBH}$ of the PBHs, \JS{$\text{erf}(X)$ is the error function, and the function $G(X)$ is given by \citep{2008gady.book.....B}:}

\begin{equation}
    G(X)=\frac{1}{2X^2}\left[\text{erf}(X)-\frac{2X}{\sqrt{\pi}}\exp{\left(-X^2\right)}\right].
\end{equation}

The velocity dispersion $\sigma_\text{PBH}$ is a direct output of {\fontfamily{qcr}\selectfont PhaseFlow} as a function of radius, whereas for the former we assume that $v=\sqrt{3}\sigma_\star=\sqrt{3}\sigma_\star(r)$, i.e. that it is given by their root-mean-square speed.  We find that the heating-to-cooling ratios are  bigger than one in all simulations but they all drop on the course of time -- sometimes drastically. In fact, the ratios drop from $\sim$$14$ ($t=0$) to $\sim$$11$ ($t=12\,\text{Gyr}$) for $R_{0,\star}=50\,\text{pc}$, from $\sim$$100$ to $\sim$$11$ for $R_{0,\star}=35\,\text{pc}$, from $\sim$$300$ to $\sim$$12$ for $R_{0,\star}=20\,\text{pc}$, and from $\sim$$800$ to $\sim$$12$ for $R_{0,\star}=10\,\text{pc}$. In all cases, the ratios appear to be constant with radius. Remarkably, all of the final cooling rates amount to almost $10$ per cent of the corresponding heating rates -- even though they were initially significantly lower. Therefore, one has to take into account the cooling due to dynamical friction and the increase of the stellar velocity dispersion, in order to simulate the dynamical evolution reliably.

Finally, the central stellar values (density and velocity dispersion) plotted in Figure~\ref{fig:mono-halfmass-evo} are fairly insensitive to the initial stellar scale radius $R_{0,\star}$ after 12~Gyr. The final half-mass radius range also shrinks, compared to the initial range, albeit less so than in the case of the central stellar quantities.

\begin{figure*}
\vspace{-5pt}
\centering
\begin{overpic}[width=1\textwidth]{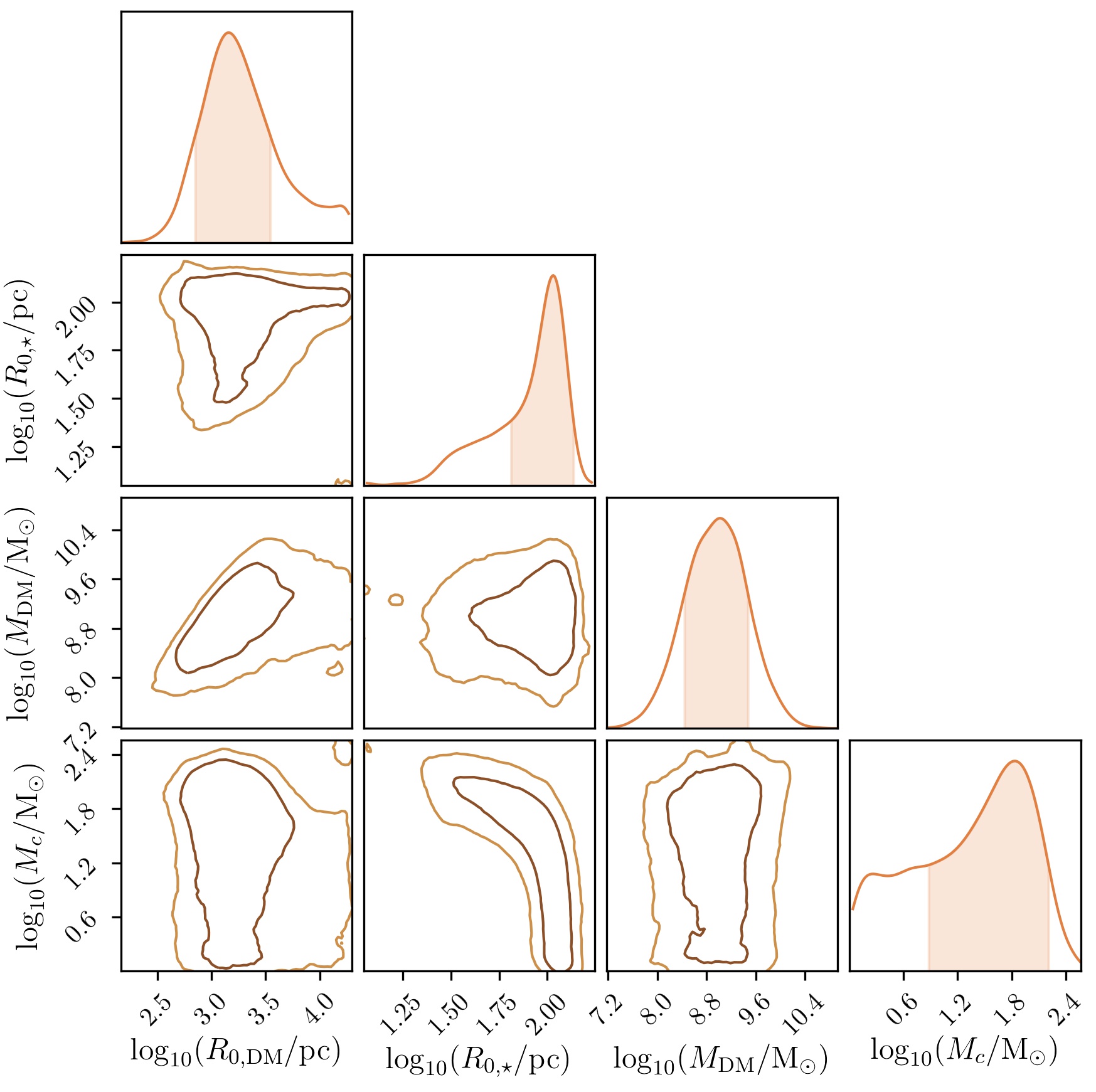}
\end{overpic}
\caption{Corner plot of the posterior distributions for $R_{0,\text{DM}}$, $R_{0,\star}$, $M_\text{DM}$, and $M_{\rm c}$, based on the stellar velocity dispersion and half-light radius of Leo IV (cf. \citealt[figure~5]{Zhu_et_al_2018}). The PBH mass function is monochromatic and the stellar mass-to-light ratio and the Coulomb logarithm are $2\,{\rm M}_\odot/{\rm L}_\odot$ and 15, respectively. Note that the prior distributions for $R_{0,\text{DM}}$, $R_{0,\star}$, and $M_{\rm c}$ are log-uniform, whereas that for $M_\text{DM}$ is log-normal. The light and dark brown contours enclose the 2$\sigma$ and 1$\sigma$ areas, respectively. Analogously, the filled areas below the curves indicate the $1\sigma$ interval with respect to their maxima.}
\vspace{-5pt}
\label{fig:monochromatic-2-15-Leo-IV}
\end{figure*}

In Figure~\ref{fig:monochromatic-2-15-Leo-IV} we show the corner plot of the posterior distributions for $R_{0,\text{DM}}$, $R_{0,\star}$, $M_\text{DM}$, and $M_{\rm c}$ resulting from the MCMC analysis on the stellar velocity dispersion and half-light radius of Leo IV (cf. \citealt{Zhu_et_al_2018}, figure~5). There, we use a stellar mass-to-light ratio of $2\,{\rm M}_\odot/{\rm L}_\odot$ and  $\ln\Lambda=15$. The posteriors for quantities which followed a log-uniform prior ($R_{0,\text{DM}}$, $R_{0,\star}$, and $M_{\rm c}$) quickly develop a pronounced peak, whereas the posterior for $M_\text{DM}$ is dominated by our choice of a log-normal prior. The light and dark brown contours enclose 2$\sigma$ and 1$\sigma$ areas, respectively. Analogously, the filled areas below the curves indicate the $1\sigma$ interval with respect to their maxima. We emphasise that the corner plot we show for Leo IV is exemplary for the other UFDGs, as their posterior distributions have a similar shape. Most importantly for this work, all  the UFDGs develop pronounced peaks in the posterior distributions for the PBH mass $M_{\rm c}$. We take the respective $1\sigma$ interval around these peaks as the PBH mass region preferred by the respective UFDG. For Leo IV, a wide range, $M_{\rm c}\in[8,162]$~M$_\odot$, is thus allowed.

\begin{landscape}
\begin{figure}
\vspace{50pt}
\centering
\begin{overpic}[width=1.35\textwidth]{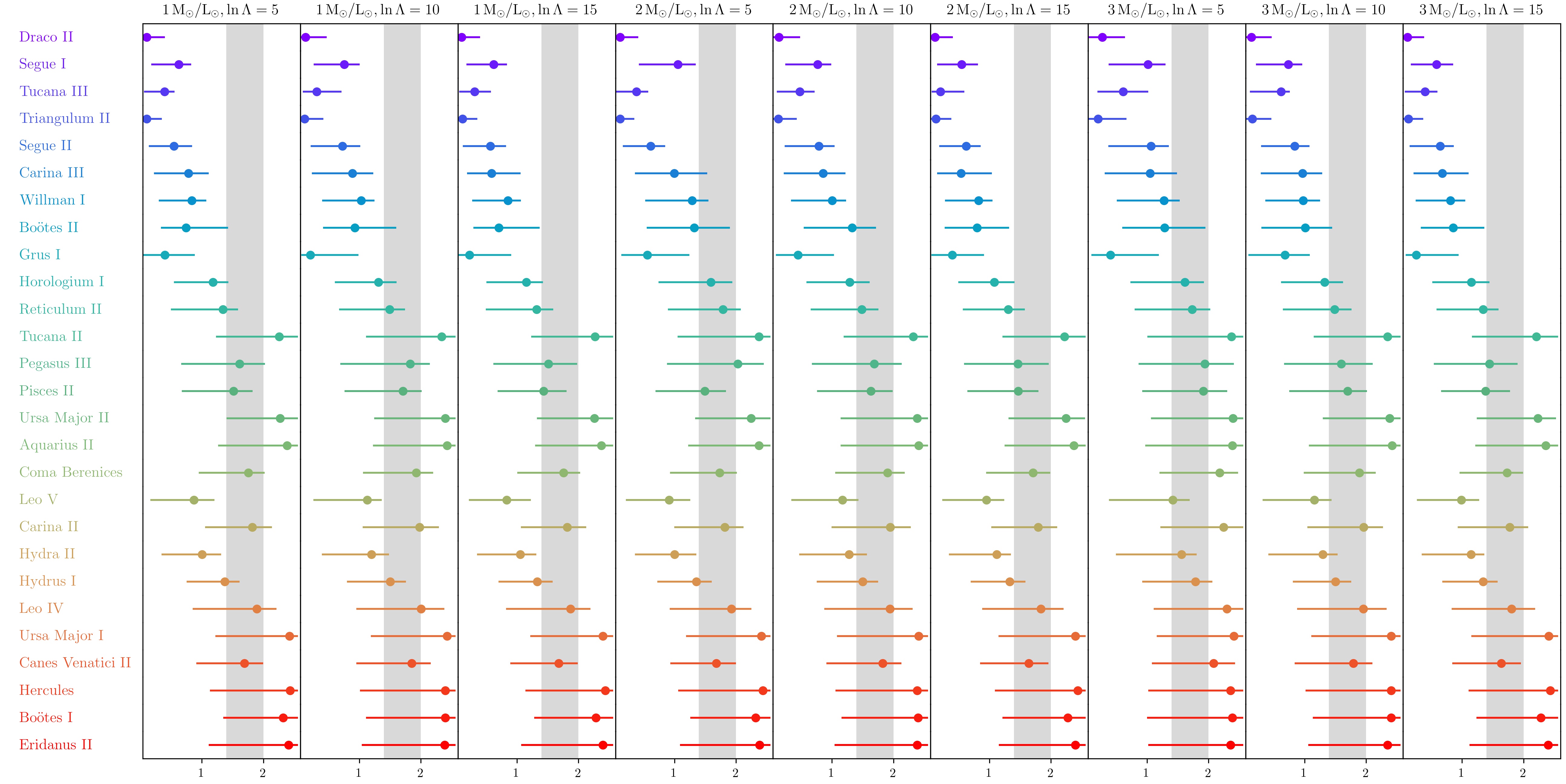}
\put (50.5,-2) {\textcolor{black}{log$_{10} (M_{\rm c}/{\rm M}_\odot)$}}
\end{overpic}
\vspace{25pt}
\caption{Preferred $1\sigma$ intervals of the monochromatic PBH mass for each individual UFDG and for different values of the stellar mass-to-light ratio (1, 2, and 3~M$_\odot/$L$_\odot$) and of the Coulomb logarithm (5, 10, and 15). There exists no single value of the PBH mass that meets all preferences simultaneously. The grey-shaded areas indicate the mass window $[25,100]$~M$_\odot$ allowed by previous constraints [emerging from the limit $\sigma\rightarrow0$ in \citet{Carr_et_al_2017}], which is incompatible with the faintest UFDGs.}
\vspace{0pt}
\label{fig:monochromatic-overview}
\end{figure}
\end{landscape}

In Figure~\ref{fig:monochromatic-overview}, we show the PBH mass regions preferred by each UFDG separately and also present the results for different values of the stellar mass-to-light ratio (1, 2, and 3~M$_\odot/$L$_\odot$) and of the Coulomb logarithm (5, 10, and 15). The UFDGs are sorted from top to bottom from the lightest (faintest) to the heaviest (brightest) galaxy; from this, a clear tendency is inferrable: {\it light (faint) UFDGs prefer light PBHs, whereas heavy (bright) UFDGs prefer heavy PBHs}. This can be concluded for all stellar mass-to-light ratios and Coulomb logarithms. In fact, the intersection of all preferred mass intervals is always empty. Put differently, {\it there exists no single value of the PBH mass that meets all preferences simultaneously}, as there are always multiple UFDGs disfavouring it. This is a key result of this paper. In the figure, we also show the mass interval $M_{\rm c}\in[25,100]$~M$_\odot$ allowed by previous constraints \citep{Carr_et_al_2017}.\footnote{Note that the monochromatic mass function we investigate at this point emerges from the log-normal mass function [Eq.~\eqref{eq:log-normal-0}] in the limit $\sigma\rightarrow 0$. In that case, $\bar{M}=\mu\exp\left(-\sigma^2/2\right)\rightarrow\mu$. Hence, the constraint on the log-normal mass function from \citet{Carr_et_al_2017} given by Eq.~\eqref{eq:mu-sigma-plane} can be translated to the monochromatic function as $M_{\rm c}\in[25,100]$~M$_\odot$.} For all choices of the stellar mass-to-light ratio and of the Coulomb logarithm, the UFDGs Draco II, Segue I, Tucana III, Triangulum II, \JS{Segue II}, and Grus I are incompatible with the previous constraints. Considering the default values (stellar mass-to-light ratio $= 2$~M$_\odot/$L$_\odot$ and $\ln\Lambda=15$), even the nine faintest UFDGs and Leo V turn out to be incompatible with that mass interval.

When using the MCMC analysis in order to infer the preferred parameters for PBH-DM, we \textit{implicitly} assume that the DM consists of PBHs, i.e. we do not question the very existence of PBHs but only the likelihood of the input parameters. In fact, given an UFDG, the inferred preferred ranges for its input parameters $R_{0,\text{DM}}$, $R_{0,\star}$, $M_\text{DM}$, and $M_{\rm c}$ do not necessarily lead to the observed stellar half-mass radius $R_{\rm h ,\star}$ and velocity dispersion $\sigma_\star$ within the reported uncertainty (cf. Table~\ref{tab:sample}). Instead, they could lead to values that are just not as bad as all the other values in the explored parameter space. Nevertheless, they might sit at the low-probability tails of Eq.~\eqref{eq:split-normal}. For instance, running a simulation (stellar mass-to-light ratio $=2$~M$_\odot/$L$_\odot$, $\ln\Lambda=15$) of Leo IV with its \textit{likeliest} input parameters, $R_{0,\text{DM}}=10^{3.16}\,\text{pc}$, $R_{0,\star}=10^{2.03}\,\text{pc}$, $M_\text{DM}=10^{9.03}\msun{}$, and $M_{\rm c}=10^{1.84}\msun{}$, i.e. the maxima of the posterior distributions in Figure~\ref{fig:monochromatic-2-15-Leo-IV} (cf. Table \ref{tab:preferences-2-15}), yields the stellar velocity dispersion $\sigma_{\rm \star,S}=3.6$~km~s$^{-1}$ and the deprojected half-mass radius $R_{\rm h,\star,S}=181.4$~pc.\footnote{As done in Section \ref{subsec:observation}, we explicitly retain the subscript \say{S} in order to indicate values as a result of a simulation -- in contrast to the use of \say{D} for the observational data.} This result has to be compared with the observed values $\sigma_{\rm \star,D}=3.3\pm1.7$~km~s$^{-1}$ and $R_{\rm h,\star,D}=(4/3)r_{\rm h,\star,D}=152\pm16$~pc (cf. Table~\ref{tab:sample}). While the simulated and observed stellar velocity dispersions agree well with each other, the proposed half-mass radius is actually too large to meet the observed one within the given measurement uncertainty. We repeat this analysis for each individual UFDG and add the results to Table \ref{tab:preferences-2-15}. For comparison, we also add the observed values adopted from Table~\ref{tab:sample}. It can be seen that the proposed stellar velocity dispersions $\sigma_{\star,\text{S}}$ of \JS{a large} majority of UFDGs \JS{(25 out of 27)} match the observed values within the respective measurement uncertainty, the exceptions being \JS{Reticulum II and Hercules, for which the simulations yield values that are slightly too large}. In contrast, the proposed half-mass radii for \JS{16} out of 27 UFDGs do not agree with the observed values within the measurement uncertainty. They are: Draco II, Segue II, Willman I, Reticulum II, Pisces II, Ursa Major II, Aquarius II, Coma Berenices, Carina II, \JS{Hydra II}, Hydrus I, Leo IV, Canes Venatici II, Hercules, Bo\"{o}tes I, and Eridanus II. The simulations of \JS{all of them except Draco II} output a value which is too large. This outcome constrains the monochromatic PBHs mass functions even more severely than the result from the MCMC analysis described in the previous paragraph. There, we concluded from the combined analysis of different UFDGs that there exists no single PBH mass that works for all of them at the same time, although we \textit{implicitly} assumed that PBHs constitute DM in each UFDG. Here, in turn, we find that the majority of UFDGs conflicts this very assumption individually since PBHs cannot reproduce their stellar half-mass radii at all. 

\begin{landscape}
\begin{table}
\vspace{40pt}
    \centering
    \caption{Preferred input parameters $R_{0,\text{DM}}$, $R_{0,\star}$, $M_\text{DM}$, and $M_{\rm c}$ for each UFDG separately. They are evaluated as the $1\sigma$ intervals around the maxima of the respective posterior distributions. For each individual UFDG, a simulation has been performed with the given input parameters yielding a deprojected stellar half-mass radius and stellar velocity dispersion, reported as $R_{\text{h},\star,\text{S}}$ and $\sigma_{\star,\text{S}}$, respectively. For comparison, we also add the observed values, $R_{\text{h},\star,\text{D}}=(4/3)r_{\text{h},\star,\text{D}}$ and $\sigma_{\star,\text{D}}$, adopted from Table~\ref{tab:sample}. The PBH mass function is monochromatic and the stellar mass-to-light ratio and the Coulomb logarithm are 2~M$_\odot/$L$_\odot$ and 15, respectively.}
    \vspace{5pt}
    \label{tab:preferences-2-15}
    \begingroup
    \renewcommand*{\arraystretch}{1.2}
    \begin{tabular}{lcccccccc}
        \hline
        \hline
		UFDG Name & $\log_{10}(R_{0,\mathrm{DM}}/\text{pc})$ & $\log_{10}(R_{0,\star}/\text{pc})$ & $\log_{10}(M_{\mathrm{DM}}/\msun{})$ & $\log_{10}(M_{\rm c}/\msun{})$ & $R_{\text{h},\star,\text{S}}$ & $\sigma_{\star,\text{S}}$ & $R_{\text{h},\star,\text{D}}$ & $\sigma_{\star,\text{D}}$\\ 
		& & & & & pc & km~s$^{-1}$ & pc & km~s$^{-1}$\\
		\hline
		Draco II & \JS{$2.92^{+0.55}_{-0.41}$} & \JS{$1.066^{+0.147}_{-0.043}$} & \JS{$9.08^{+0.55}_{-0.47}$} & \JS{$0.115^{+0.291}_{-0.075}$} & \JS{$21.4$} & \JS{$1.2$} & $25.3^{+6.0}_{-3.5}$ & $<5.9$ ($95\%$ C.L.)\\ 
		Segue I & $2.72^{+0.27}_{-0.28}$ & $1.294^{+0.090}_{-0.182}$ & $9.00^{+0.56}_{-0.45}$ & $0.55^{+0.26}_{-0.40}$ & $35.3$ & $3.2$ & $32.3\pm3.7$ & $3.7^{+1.4}_{-1.1}$\\
		Tucana III & \JS{$3.53^{+0.47}_{-0.30}$} & \JS{$1.39^{+0.13}_{-0.22}$} & \JS{$9.08^{+0.42}_{-0.56}$} & \JS{$0.20^{+0.39}_{-0.14}$} & \JS{$42.6$} & \JS{$0.3$} & $45\pm11$ & $<1.2$ ($90\%$ C.L.)\\ 
		Triangulum II & \JS{$2.96^{+0.64}_{-0.33}$} & \JS{$1.065^{+0.129}_{-0.043}$} & \JS{$9.17^{+0.46}_{-0.56}$} & \JS{$0.128^{+0.254}_{-0.088}$} & \JS{$22.4$} & \JS{$1.2$} & $23.2\pm5.7$ & $<3.4$ ($90\%$ C.L.)\\ 
		Segue II & \JS{$3.37^{+0.56}_{-0.30}$} & \JS{$1.49^{+0.10}_{-0.23}$} & \JS{$8.94^{+0.52}_{-0.51}$} & \JS{$0.62^{+0.24}_{-0.44}$} & \JS{$57.1$} & \JS{$0.6$} & $51.1\pm3.7$ & $<2.6$ ($95\%$ C.L.)\\ 
		Carina III & $2.67^{+0.35}_{-0.34}$ & $1.31\pm 0.19$ & $8.94^{+0.63}_{-0.42}$ & $0.54^{+0.50}_{-0.39}$ & $35.0$ & $3.5$ & $40\pm12$ & $5.6^{+4.3}_{-2.1}$\\ 
		Willman I & $2.70\pm 0.28$ & $1.36^{+0.11}_{-0.22}$ & $9.16^{+0.37}_{-0.66}$ & $0.82^{+0.22}_{-0.55}$ & $41.0$ & $4.8$ & $36.9\pm3.2$ & $4.0\pm0.8$\\ 
		Bo\"{o}tes II & $2.54^{+0.55}_{-0.39}$ & $1.54^{+0.11}_{-0.23}$ & $9.02^{+0.46}_{-0.52}$ & $0.80\pm 0.52$ & $49.8$ & $7.8$ & $51.6\pm6.8$ & $10.5\pm7.4$\\ 
		Grus I & $3.00^{+0.74}_{-0.44}$ & $1.26^{+0.27}_{-0.19}$ & $8.97^{+0.51}_{-0.52}$ & $0.39^{+0.52}_{-0.34}$ & $32.2$ & $1.3$ & $37.7\pm30.7$ & $2.9^{+6.9}_{-2.1}$\\
		Horologium I & $2.69^{+0.23}_{-0.24}$ & $1.51^{+0.11}_{-0.27}$ & $9.02^{+0.50}_{-0.54}$ & $1.08^{+0.33}_{-0.58}$ & $54.3$ & $5.6$ & $48.7\pm9.5$ & $4.9^{+2.8}_{-0.9}$\\ 
		Reticulum II & $2.93\pm 0.32$ & $1.64^{+0.11}_{-0.39}$ & $9.21^{+0.29}_{-0.71}$ & $1.30^{+0.27}_{-0.73}$ & $77.2$ & $4.3$ & $64.3\pm2.3$ & $3.3\pm0.7$\\ 
		Tucana II & $2.80^{+0.30}_{-0.27}$ & $2.04^{+0.19}_{-0.36}$ & $8.95^{+0.47}_{-0.52}$ & $2.22^{+0.34}_{-1.01}$ & $196.2$ & $9.6$ & $160\pm40$ & $8.6^{+4.4}_{-2.7}$\\ 
		Pegasus III & $2.88^{+0.40}_{-0.34}$ & $1.81^{+0.19}_{-0.36}$ & $9.02^{+0.46}_{-0.55}$ & $1.46^{+0.50}_{-0.87}$ & $105.8$ & $5.2$ & $104^{+40}_{-32}$ & $5.4^{+3.0}_{-2.5}$\\ 
		Pisces II & $2.84^{+0.37}_{-0.36}$ & $1.74^{+0.12}_{-0.26}$ & $8.91^{+0.57}_{-0.44}$ & $1.47^{+0.33}_{-0.83}$ & $98.0$ & $4.8$ & $79.1\pm11.3$ & $5.4^{+3.6}_{-2.4}$\\ 
		Ursa Major II & $3.01^{+0.23}_{-0.25}$ & $2.101^{+0.097}_{-0.284}$ & $9.06^{+0.43}_{-0.58}$ & $2.25^{+0.31}_{-0.94}$ & $213.2$ & $6.7$ & $170.7\pm6.7$ & $5.6\pm1.4$\\ 
		Aquarius II & $3.01^{+0.22}_{-0.23}$ & $2.19^{+0.14}_{-0.26}$ & $8.98^{+0.46}_{-0.54}$ & $2.37^{+0.19}_{-1.13}$ & $263.8$ & $6.9$ & $212\pm32$ & $5.4^{+3.4}_{-0.9}$\\ 
		Coma Berenices & $2.91^{+0.20}_{-0.24}$ & $1.842^{+0.086}_{-0.277}$ & $8.97^{+0.49}_{-0.53}$ & $1.71^{+0.28}_{-0.76}$ & $121.0$ & $5.2$ & $96.1\pm5.1$ & $4.6\pm0.8$\\ 
		Leo V & $3.11^{+0.56}_{-0.43}$ & $1.64^{+0.15}_{-0.31}$ & $8.90^{+0.58}_{-0.46}$ & $0.95^{+0.29}_{-0.72}$ & $75.8$ & $1.7$ & $69.1\pm22.1$ & $2.3^{+3.2}_{-1.6}$\\
		Carina II & $3.07^{+0.22}_{-0.24}$ & $1.938^{+0.091}_{-0.285}$ & $9.00^{+0.49}_{-0.54}$ & $1.79^{+0.31}_{-0.76}$ & $148.9$ & $3.9$ & $121\pm11$ & $3.4^{+1.2}_{-0.8}$\\ 
		Hydra II & \JS{$3.29^{+0.75}_{-0.19}$} & \JS{$1.73^{+0.12}_{-0.18}$} & \JS{$8.96^{+0.46}_{-0.53}$} & \JS{$1.12^{+0.23}_{-0.77}$} & \JS{$96.9$} & \JS{$1.3$} & $78.9\pm14.5$ & $<3.6$ ($90\%$ C.L.)\\ 
		Hydrus I & $3.02^{+0.20}_{-0.23}$ & $1.687^{+0.090}_{-0.274}$ & $9.07^{+0.43}_{-0.57}$ & $1.33^{+0.25}_{-0.64}$ & $84.5$ & $3.2$ & $71\pm5$ & $2.7\pm0.5$\\ 
		Leo IV & $3.16^{+0.39}_{-0.31}$ & $2.03^{+0.11}_{-0.22}$ & $9.03^{+0.45}_{-0.59}$ & $1.84^{+0.37}_{-0.96}$ & $181.4$ & $3.6$ & $152\pm16$ & $3.3\pm1.7$\\ 
		Ursa Major I & \JS{$3.14^{+0.19}_{-0.28}$} & \JS{$2.451^{+0.013}_{-0.061}$} & \JS{$9.06^{+0.52}_{-0.59}$} & \JS{$2.40^{+0.16}_{-1.25}$} & \JS{$395.3$} & \JS{$6.9$} & \JS{$391.2^{+45.9}_{-43.5}$} & $7.0\pm1.0$\\ 
		Canes Venatici II & $2.90\pm 0.23$ & $1.81^{+0.13}_{-0.28}$ & $8.98^{+0.50}_{-0.52}$ & $1.64^{+0.32}_{-0.79}$ & $115.4$ & $5.1$ & $94.3\pm14.9$ & $4.6\pm1.0$\\ 
		Hercules & $3.13\pm 0.21$ & $2.331^{+0.080}_{-0.110}$ & $9.03^{+0.42}_{-0.58}$ & $2.44^{+0.13}_{-1.35}$ & $320.9$ & $6.2$ & $288\pm23$ & $5.1\pm0.9$\\ 
		Bo\"{o}tes I & $3.37^{+0.22}_{-0.20}$ & $2.279^{+0.066}_{-0.174}$ & $9.07^{+0.41}_{-0.57}$ & $2.27^{+0.29}_{-1.06}$ & $326.6$ & $3.1$ & $255\pm7$ & $2.4^{+0.9}_{-0.5}$\\ 
		Eridanus II & $3.18^{+0.16}_{-0.31}$ & $2.452^{+0.019}_{-0.039}$ & $9.09^{+0.49}_{-0.60}$ & $2.40^{+0.16}_{-1.25}$ & $402.7$ & $6.5$ & $369\pm19$ & $6.9^{+1.2}_{-0.9}$\\ 
		\hline
		\hline
    \end{tabular}
    \endgroup
\end{table}
\end{landscape}

\subsection{Log-normal mass function}

Here, we investigate whether the constraints resulting from the MCMC with an extended, log-normal mass function for PBHs weaken those from  the previous section. We restrict our analysis to a stellar mass-to-light ratio of $2\msun/{\text{L}}_\odot$ and a Coulomb logarithm of 15. 

\begin{figure*}
	\centering
	\begin{overpic}[width=1.0\textwidth]{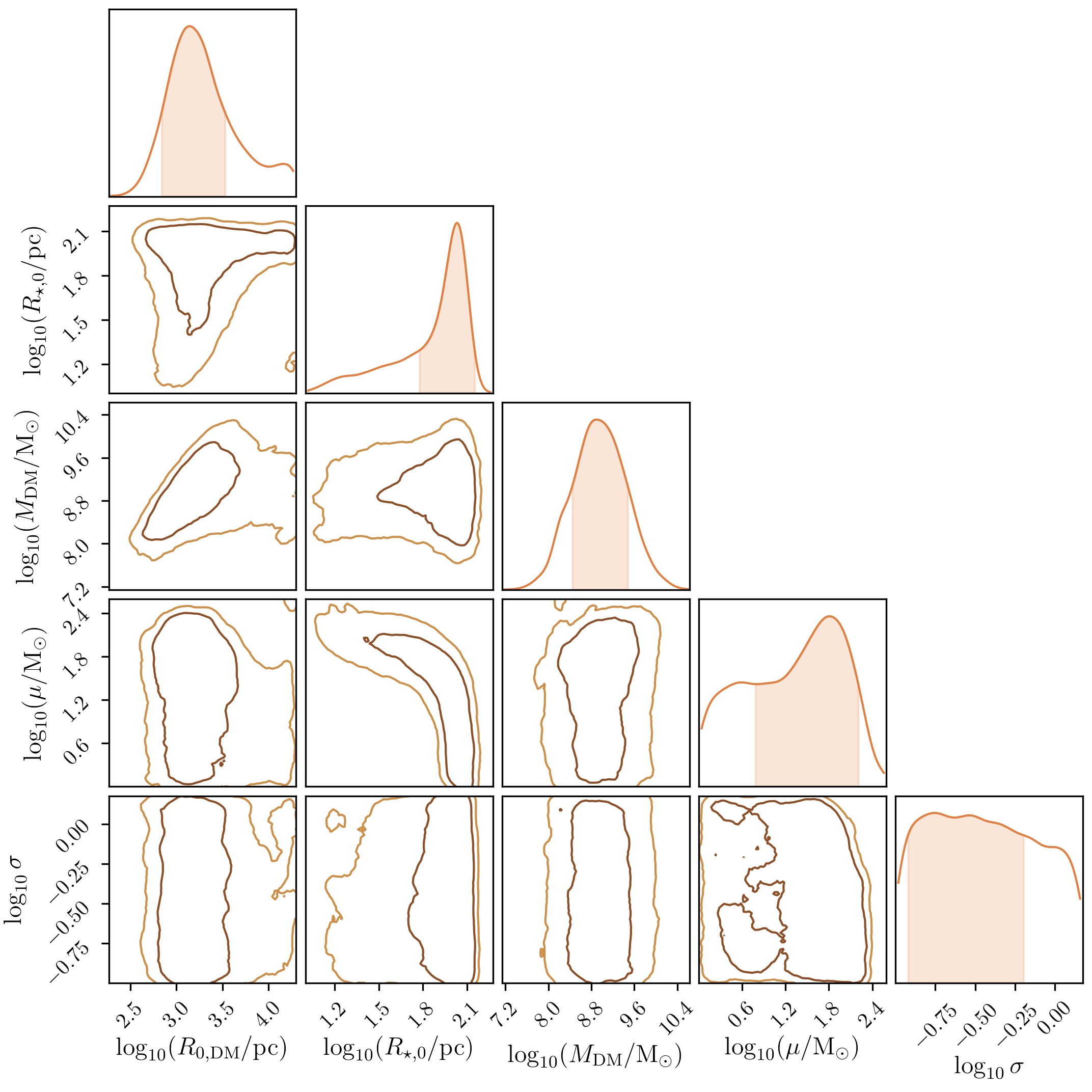}
	\end{overpic}
	\caption{Corner plot of the posterior distributions for $R_{0,\text{DM}}$, $R_{0,\star}$, $M_\text{DM}$, $\mu$, and $\sigma$ based on the stellar velocity dispersion and half-mass radius of Leo IV (cf. \citealt[figure~5]{Zhu_et_al_2018}). The PBH mass function is log-normal and the stellar mass-to-light ratio and the Coulomb logarithm are $2\msun/{\text{L}}_\odot$ and 15, respectively. Note that the prior distributions for $R_{0,\text{DM}}$, $R_{0,\star}$, $\mu$, and $\sigma$ are log-uniform, whereas that for $M_\text{DM}$ is log-normal. The light and dark orange contours enclose the 2$\sigma$ and 1$\sigma$ areas, respectively. Analogously, the filled areas below the curves indicate the $1\sigma$ interval with respect to their maxima.}
	\vspace{-5pt}
	\label{fig:extended-2-15-Leo-IV}
\end{figure*}

In Figure~\ref{fig:extended-2-15-Leo-IV}, we present the resulting corner plot for Leo IV as a representative for the others. Note that we now deal with five input parameters, $R_{0,\text{DM}}$, $R_{0,\star}$, $M_\text{DM}$, $\mu$, and $\sigma$, instead of four since the log-normal mass function depends on two parameters. Notably, a comparison with the posterior distributions for $R_{0,\text{DM}}$, $R_{0,\star}$, and $M_\text{DM}$ that resulted from the monochromatic simulations (Figure~\ref{fig:monochromatic-2-15-Leo-IV}, Table~\ref{tab:preferences-2-15}) shows no significant difference. In fact, the 1$\sigma$ intervals around the respective maxima of the posterior distributions which are listed in Table~\ref{tab:preferences-2-15-extended} agree well with the previous one: $\log_{10}(R_{0,\text{DM}}/\text{pc})=3.14^{+0.39}_{-0.30}$ ($3.16^{+0.39}_{-0.31}$ for the monochromatic case), $\log_{10}(R_{0,\star}/\text{pc})=2.03^{+0.12}_{-0.26}$ ($2.03^{+0.11}_{-0.22}$), and $\log_{10}(M_\text{DM}/\msun)=8.90^{+0.58}_{-0.45}$ ($9.03^{+0.45}_{-0.59}$). Concerning the posterior distribution for $\mu$, we have to bear in mind that it is, in general, not directly comparable to $M_{\rm c}$, since they have different meanings. The latter is equivalent to the mean mass of the monochromatic mass distribution, whereas this is given by $\bar{M}=\mu\exp(-\sigma^2/2)$ for the log-normal. At this point, we refer to \citet{2016PhRvD..94f3530G}, who proposed to obtain a general constraint on an extended mass function by a replacement that, in our case ($\ln\Lambda=15$ and $f_\text{PBH}=1$), reads:

\begin{equation}\label{eq:new-repl}
M_{\rm c}\rightsquigarrow\int_{0}^{\infty}\psi(M)M\mathrm{d}M=\mu\exp\left(+\frac{\sigma^2}{2}\right).
\end{equation}

For Leo IV, $\log_{10}\sigma=-0.75^{+0.56}_{-0.17}$ is small, meaning that $\mu\exp(+\sigma^2/2)\approx\mu$. Indeed, $\log_{10}(\mu/\msun)=1.81^{+0.39}_{-1.03}$ agrees well with $\log_{10}(M_{\rm c}/\msun)=1.84^{+0.37}_{-0.96}$, confirming the proposed method by \citet{2016PhRvD..94f3530G}.

\begin{figure*}
	\centering
	\begin{overpic}[width=1.0\textwidth]{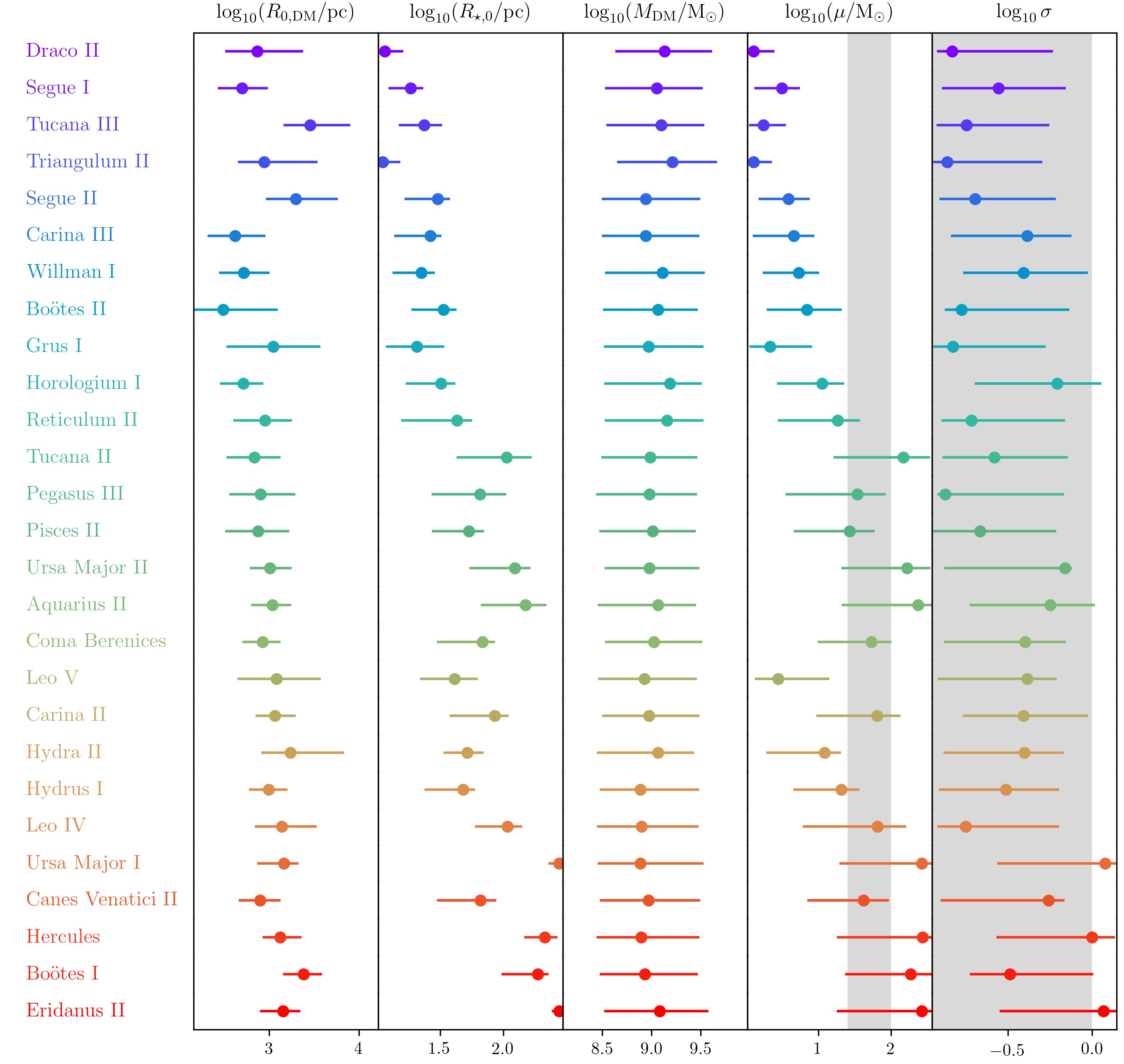}
	\end{overpic}
	\caption{Preferred $1\sigma$ intervals of the input parameters for the log-normal PBH mass function. The stellar mass-to-light ratio and the Coulomb logarithm are 2 M$_\odot/$L$_\odot$ and 15, respectively. The fourth column ($\log_{10}\mu$) has to be compared with the corresponding column (2 M$_\odot/$L$_\odot$, $\ln\Lambda=15$) of Figure~\ref{fig:monochromatic-overview}. Still, there exists no single value of the PBH mass that meets all preferences simultaneously since the stellar kinematics appear to be fairly insensitive to the width $\sigma$ of the mass function. The grey-shaded areas indicate the $\mu$-window $[25,100]$~M$_\odot$ and $\sigma$-window $[0.0,1.0]$ allowed by previous constraints \citep{Carr_et_al_2017}.}
	\vspace{0pt}
	\label{fig:extended-overview}
\end{figure*}

In Figure~\ref{fig:extended-overview}, we show the preferred input parameters for all UFDGs separately and add them to Table~\ref{tab:preferences-2-15-extended}. We find no value of $\mu$ that is allowed by all UFDGs, analogously to what found in Section \ref{sec:mono-results}. Again, there is the tendency for \textit{light (faint) UFDGs to prefer light PBHs and heavy (bright) UFDGs to prefer heavy PBHs}. The nine lightest UFDGs still constitute the critical sub-sample that by itself would conflict with the previous constraints \citep{Carr_et_al_2017}. In turn, the constraints of \citet{Carr_et_al_2017} agree well with the preferred ranges for $\sigma$ of all our UFDGs. In fact, there is much scatter across the allowed range, suggesting that the constraint is fairly insensitive to the actual value of $\sigma$. If the result  depends on $\mu\exp(+\sigma^2/2)$, as suggested by Eq. \eqref{eq:new-repl}, this is reasonable as the contribution of $\sigma$ would be exponentially suppressed for all UFGDs.

Finally, we also investigate whether the \textit{likeliest} values for the input parameters can be used to reconstruct the observed properties of the stellar populations. For this purpose, we proceed in the same way as in Section \ref{sec:mono-results} and add the resulting stellar half-mass radii $R_{\text{h},\star,{\rm S}}$ and stellar velocity dispersions $\sigma_{\star,{\rm S}}$ to Table~\ref{tab:preferences-2-15-extended}. \JS{Again, the observed stellar velocity dispersions $\sigma_{\star,{\rm D}}$ can be well reproduced. Indeed, \textit{all} UFDGs agree within the given measurement uncertainty.} In turn, the half-mass radii of \JS{16} UFDGs are incompatible with the observed, $R_{\text{h},\star,{\rm D}}$. 

From the  analysis above, we conclude that the log-normal mass function does not weaken the constraints we found for the monochromatic mass function. In fact, the width of the distribution seems to have little or no effect on the dynamical constraints.

\begin{landscape}
	\begin{table}
		\centering
		\caption{The \textit{likeliest} input parameters $R_{0,\text{DM}}$, $R_{0,\star}$, $M_\text{DM}$, $\mu$, and $\sigma$ from the MCMC with a log-normal mass function (cf. Table~\ref{tab:preferences-2-15} for the monochromatic case). They are evaluated as the $1\sigma$ intervals around the maxima of the respective posterior distributions. Again, we performed for each individual UFDG a simulation with these values resulting in the listed values for $R_{\text{h},\star,{\rm S}}$ and $\sigma_{\star,{\rm S}}$. For comparison, we also add the observed values, $R_{\text{h},\star,{\rm D}}=(4/3)r_{\text{h},\star,{\rm D}}$ and $\sigma_{\star,{\rm D}}$, adopted from Table~\ref{tab:sample}. The stellar mass-to-light ratio and the Coulomb logarithm were 2~M$_\odot/$L$_\odot$ and 15, respectively.}
		\label{tab:preferences-2-15-extended}
		\begingroup
		\renewcommand*{\arraystretch}{1.2}
		\begin{tabular}{lccccccccc}
			\hline
			\hline
			UFDG Name & $\log_{10}(R_{0,\mathrm{DM}}/\text{pc})$ & $\log_{10}(R_{0,\star}/\text{pc})$ & $\log_{10}(M_{\mathrm{DM}}/\msun{})$ & $\log_{10}(\mu/\msun{})$ & $\log_{10}\sigma$ & $R_{\text{h},\star,{\rm S}}$ & $\sigma_{\star,{\rm S}}$ & $R_{\text{h},\star,{\rm D}}$ & $\sigma_{\star,{\rm D}}$\\ 
			& & & & & & pc & km~s$^{-1}$ & pc & km~s$^{-1}$\\
			\hline
			Draco II & \JS{$2.87^{+0.51}_{-0.36}$} & \JS{$1.062^{+0.148}_{-0.041}$} & \JS{$9.13^{+0.48}_{-0.50}$} & \JS{$0.101^{+0.292}_{-0.062}$} & \JS{$-0.833^{+0.599}_{-0.089}$} & \JS{$20.5$} & \JS{$1.5$} & $25.3^{+6.0}_{-3.5}$ & $<5.9$ ($95\%$ C.L.)\\ 
			Segue I & $2.70^{+0.29}_{-0.27}$ & $1.27^{+0.10}_{-0.17}$ & $9.05^{+0.47}_{-0.53}$ & $0.49^{+0.25}_{-0.38}$ & $-0.56^{+0.40}_{-0.34}$ & $31.5$ & $3.4$ & $32.3\pm3.7$ & $3.7^{+1.4}_{-1.1}$\\ 
			Tucana III & \JS{$3.46^{+0.45}_{-0.30}$} & \JS{$1.38^{+0.14}_{-0.20}$} & \JS{$9.10^{+0.44}_{-0.56}$} & \JS{$0.24^{+0.31}_{-0.20}$} & \JS{$-0.75^{+0.49}_{-0.18}$} & \JS{$42.0$} & \JS{$0.4$} & $45\pm11$ & $<1.2$ ($90\%$ C.L.)\\ 
			Triangulum II & \JS{$2.94^{+0.60}_{-0.29}$} & \JS{$1.046^{+0.139}_{-0.028}$} & \JS{$9.21^{+0.46}_{-0.56}$} & \JS{$0.103^{+0.252}_{-0.063}$} & \JS{$-0.861^{+0.565}_{-0.084}$} & \JS{$20.6$} & \JS{$1.3$} & $23.2\pm5.7$ & $<3.4$ ($90\%$ C.L.)\\ 
			Segue II & \JS{$3.30^{+0.47}_{-0.33}$} & \JS{$1.482^{+0.097}_{-0.262}$} & \JS{$8.94^{+0.55}_{-0.45}$} & \JS{$0.58^{+0.30}_{-0.41}$} & \JS{$-0.70^{+0.48}_{-0.21}$} & \JS{$56.0$} & \JS{$0.7$} & $51.1\pm3.7$ & $<2.6$ ($95\%$ C.L.)\\ 
			Carina III & $2.62^{+0.34}_{-0.31}$ & $1.424^{+0.088}_{-0.286}$ & $8.94^{+0.55}_{-0.44}$ & $0.66^{+0.28}_{-0.56}$ & $-0.39^{+0.26}_{-0.45}$ & $41.3$ & $4.9$ & $40\pm12$ & $5.6^{+4.3}_{-2.1}$\\ 
			Willman I & $2.72^{+0.29}_{-0.28}$ & $1.35^{+0.11}_{-0.23}$ & $9.11^{+0.43}_{-0.59}$ & $0.72^{+0.29}_{-0.50}$ & $-0.41^{+0.38}_{-0.36}$ & $37.7$ & $4.2$ & $36.9\pm3.2$ & $4.0\pm0.8$\\ 
			Bo\"{o}tes II & $2.48^{+0.61}_{-0.34}$ & $1.53^{+0.10}_{-0.25}$ & $9.07^{+0.40}_{-0.56}$ & $0.84^{+0.48}_{-0.56}$ & $-0.78^{+0.64}_{-0.10}$ & $49.5$ & $9.4$ & $51.6\pm6.8$ & $10.5\pm7.4$\\ 
			Grus I & $3.05^{+0.53}_{-0.52}$ & $1.32^{+0.22}_{-0.25}$ & $8.97^{+0.56}_{-0.45}$ & $0.33^{+0.59}_{-0.28}$ & $-0.83^{+0.55}_{-0.12}$ & $35.0$ & $1.2$ & $37.7\pm30.7$ & $2.9^{+6.9}_{-2.1}$\\ 
			Horologium I & $2.71^{+0.22}_{-0.26}$ & $1.51^{+0.11}_{-0.28}$ & $9.19^{+0.32}_{-0.67}$ & $1.05^{+0.30}_{-0.62}$ & $-0.21^{+0.26}_{-0.49}$ & $53.6$ & $6.3$ & $48.7\pm9.5$ & $4.9^{+2.8}_{-0.9}$\\ 
			Reticulum II & $2.95^{+0.30}_{-0.35}$ & $1.63^{+0.12}_{-0.44}$ & $9.16^{+0.37}_{-0.63}$ & $1.26^{+0.31}_{-0.83}$ & $-0.72^{+0.56}_{-0.18}$ & $77.0$ & $3.8$ & $64.3\pm2.3$ & $3.3\pm0.7$\\ 
			Tucana II & $2.83^{+0.29}_{-0.31}$ & $2.03^{+0.20}_{-0.40}$ & $8.99^{+0.48}_{-0.50}$ & $2.17^{+0.37}_{-0.97}$ & $-0.58^{+0.44}_{-0.31}$ & $188.4$ & $8.9$ & $160\pm40$ & $8.6^{+4.4}_{-2.7}$\\ 
			Pegasus III & $2.90^{+0.39}_{-0.35}$ & $1.82^{+0.21}_{-0.38}$ & $8.98^{+0.48}_{-0.54}$ & $1.54^{+0.39}_{-0.99}$ & $-0.873^{+0.707}_{-0.048}$ & $114.2$ & $4.9$ & $104^{+40}_{-32}$ & $5.4^{+3.0}_{-2.5}$\\ 
			Pisces II & $2.88^{+0.35}_{-0.37}$ & $1.73^{+0.12}_{-0.29}$ & $9.01^{+0.44}_{-0.54}$ & $1.43^{+0.34}_{-0.77}$ & $-0.67^{+0.45}_{-0.28}$ & $93.6$ & $4.8$ & $79.1\pm11.3$ & $5.4^{+3.6}_{-2.4}$\\ 
			Ursa Major II & $3.01^{+0.24}_{-0.23}$ & $2.09^{+0.12}_{-0.36}$ & $8.98^{+0.51}_{-0.45}$ & $2.22^{+0.32}_{-0.91}$ & $-0.162^{+0.042}_{-0.716}$ & $231.5$ & $6.4$ & $170.7\pm6.7$ & $5.6\pm1.4$\\ 
			Aquarius II & $3.04^{+0.21}_{-0.24}$ & $2.18^{+0.17}_{-0.35}$ & $9.06^{+0.39}_{-0.61}$ & $2.37^{+0.19}_{-1.05}$ & $-0.25^{+0.27}_{-0.48}$ & $264.9$ & $7.2$ & $212\pm32$ & $5.4^{+3.4}_{-0.9}$\\ 
			Coma Berenices & $2.92^{+0.21}_{-0.22}$ & $1.84^{+0.10}_{-0.36}$ & $9.02\pm 0.49$ & $1.73^{+0.27}_{-0.74}$ & $-0.40^{+0.24}_{-0.48}$ & $122.4$ & $5.4$ & $96.1\pm5.1$ & $4.6\pm0.8$\\ 
			Leo V & $3.08^{+0.49}_{-0.43}$ & $1.62^{+0.18}_{-0.27}$ & $8.93^{+0.53}_{-0.47}$ & $0.44^{+0.71}_{-0.32}$ & $-0.39^{+0.18}_{-0.53}$ & $60.8$ & $1.5$ & $69.1\pm22.1$ & $2.3^{+3.2}_{-1.6}$\\ 
			Carina II & $3.06^{+0.24}_{-0.21}$ & $1.93^{+0.11}_{-0.36}$ & $8.97^{+0.51}_{-0.47}$ & $1.81^{+0.32}_{-0.84}$ & $-0.41^{+0.38}_{-0.36}$ & $155.1$ & $4.0$ & $121\pm11$ & $3.4^{+1.2}_{-0.8}$\\ 
			Hydra II & \JS{$3.24^{+0.60}_{-0.32}$} & \JS{$1.72^{+0.13}_{-0.19}$} & \JS{$9.07^{+0.37}_{-0.62}$} & \JS{$1.08^{+0.23}_{-0.80}$} & \JS{$-0.40^{+0.24}_{-0.48}$} & \JS{$87.7$} & \JS{$1.6$} & $78.9\pm14.5$ & $<3.6$ ($90\%$ C.L.)\\ 
			Hydrus I & $2.99^{+0.21}_{-0.22}$ & $1.684^{+0.092}_{-0.306}$ & $8.89^{+0.60}_{-0.41}$ & $1.31^{+0.25}_{-0.66}$ & $-0.51^{+0.31}_{-0.40}$ & $90.7$ & $2.9$ & $71\pm5$ & $2.7\pm0.5$\\ 
			Leo IV & $3.14^{+0.39}_{-0.30}$ & $2.03^{+0.12}_{-0.26}$ & $8.90^{+0.58}_{-0.45}$ & $1.81^{+0.39}_{-1.03}$ & $-0.75^{+0.56}_{-0.17}$ & $187.7$ & $3.3$ & $152\pm16$ & $3.3\pm1.7$\\ 
			Ursa Major I & \JS{$3.16^{+0.17}_{-0.30}$} & \JS{$2.442^{+0.017}_{-0.084}$} & \JS{$8.89^{+0.64}_{-0.44}$} & \JS{$2.42^{+0.14}_{-1.14}$} & \JS{$0.078^{+0.080}_{-0.643}$} & \JS{$485.8$} & \JS{$6.1$} & \JS{$391.2^{+45.9}_{-43.5}$} & $7.0\pm1.0$\\ 
			Canes Venatici II & $2.90^{+0.23}_{-0.24}$ & $1.82^{+0.13}_{-0.34}$ & $8.97^{+0.53}_{-0.49}$ & $1.62^{+0.35}_{-0.78}$ & $-0.260^{+0.095}_{-0.640}$ & $118.6$ & $5.2$ & $94.3\pm14.9$ & $4.6\pm1.0$\\ 
			Hercules & $3.12^{+0.24}_{-0.20}$ & $2.33^{+0.10}_{-0.16}$ & $8.89^{+0.59}_{-0.45}$ & $2.44^{+0.13}_{-1.18}$ & $0.00^{+0.14}_{-0.57}$ & $384.0$ & $6.0$ & $288\pm23$ & $5.1\pm0.9$\\ 
			Bo\"{o}tes I & $3.38^{+0.21}_{-0.23}$ & $2.273^{+0.085}_{-0.284}$ & $8.93^{+0.54}_{-0.46}$ & $2.27^{+0.29}_{-0.91}$ & $-0.49^{+0.50}_{-0.24}$ & $324.9$ & $2.7$ & $255\pm7$ & $2.4^{+0.9}_{-0.5}$\\ 
			Eridanus II & $3.15^{+0.19}_{-0.26}$ & $2.441^{+0.022}_{-0.058}$ & $9.08^{+0.50}_{-0.56}$ & $2.42^{+0.14}_{-1.17}$ & $0.067^{+0.092}_{-0.616}$ & $397.2$ & $7.2$ & $369\pm19$ & $6.9^{+1.2}_{-0.9}$\\  
			\hline
			\hline
		\end{tabular}
		\endgroup
	\end{table}
\end{landscape}


\section{Discussion and Conclusions}\label{sec:discussion}

In this work, we investigated the dynamical effect of $\mathcal{O}(1$--$100)\msun$ PBH-DM on the stellar population of UFDGs by means of Fokker--Planck simulations. We showed that  PBHs heat up their stellar system by means of two-body scatterings that result in an increase of the stellar velocity dispersion and in the growth of the stellar half-mass radius. We inferred the posterior probability distributions for certain input parameters to reconstruct the observed  stellar velocity dispersions and half-mass radii of UFDGs by means of $2\times10^5$ MCMC runs per galaxy. In particular, we investigated the effect of different PBH mass functions (monochromatic and log-normal) on  the dynamical evolution of stars  inhabiting  DM-dominated UFDGs.

As a result, we are able to make three statements each of which independently challenges the PBH-DM scenario:

\begin{enumerate}[label=(\roman*),leftmargin=.4in]
	
	\item Concerning the monochromatic mass function, we found that the lowest-mass (faintest) UFDGs constitute a critical sub-sample that prefers a PBH mass range $M_{\rm c}\sim\mathcal{O}(1)\msun$; this range is compatible with some previous dynamical constraints \citep{2016ApJ...824L..31B,Zhu_et_al_2018} but incompatible with  other constraints \citep{Carr_et_al_2017}. Assuming a log-normal PBH mass function does not resolve this conflict.
	
	\item Even disregarding the previous point, the combined analysis of all 27 UFDGs in our sample leaves no open mass window for $M_{\rm c}$ (the mass for the monochromatic mass function) and $\mu$ (the scale mass for the log-normal mass function) in which PBH-DM could simultaneously produce the observed stellar properties for all galaxies.
	
	\item In a majority of UFDGs, PBH-DM would puff up the stellar systems to half-mass radii that are  too large to meet the observed values (even accounting for the uncertainty in the observables).
	
\end{enumerate}

\JS{In the following,} we discuss the possible shortcomings of our work.

First of all, the stellar observables of several among the lightest UFDGs suffer from poor statistics (as mentioned in Table~\ref{tab:sample}) due to the small number of detectable stars. This could in principle weaken our point (i) above; however, we stress that it would be extremely unlikely that all the stellar observables in this sub-sample would shift our results in the same direction. \JS{Similarly, we note that the very nature of some of the lightest UFDGs in our sample is less clear, e.g. that of Tucana III \citep{2017ApJ...838...11S} and Triangulum II \citep{2017ApJ...838...83K}. To the present, it cannot be ruled out that those are actually baryon-dominated star clusters instead of DM-dominated UFDGs. If future observations corroborate this scenario, the respective systems must not be included in our sample in order to draw meaningful conclusions on PBHs as DM. Our results can be easily updated to such situation by simply removing the respective rows in Tables \ref{tab:preferences-2-15} and \ref{tab:preferences-2-15-extended} and Figures \ref{fig:monochromatic-overview} and \ref{fig:extended-overview}. However, we highlight that none of our three main findings would change unless a majority of UFDGs in the critical sub-sample turns out to be spurious.}

Furthermore, we \JS{emphasise} that we only studied the scenario in which PBHs constitute \textit{all} of the DM (\say{PBH-DM}). \JS{For models in which DM consists only partially of PBHs, the three statements might no longer be valid.} In particular, our method is inadequate to infer any constraint on the possibility that DM is complementarily composed of PBHs and a collisionless component, as our methodology does not allow us to implement the latter. 

We additionally neglected the possible influence of gas and a central, \JS{intermediate-mass} black hole on the dynamics of UFDGs. The former assumption is well-justified as the observed UFDGs show no significant gas content [cf. \citet{2019arXiv190105465S}]. On the other hand, a massive black hole could inhabit the core of at least some among our galaxies (see, e.g. the  recent observation by \citealt{2017ApJ...850..196B} of a more massive dwarf). However, the  mass of a central black hole in such low-mass systems is expected to be very small ($\lesssim 10^4 \msun{}$) if we extrapolate the commonly accepted scaling relations \citep{Shankar_et_al_2016} to the lowest mass end. Such black holes are expected to have a non-negligible dynamical influence only at very small scales ($\lesssim 1$~pc), where a small cusp could develop \citep{Bahcall1976}. On the other hand, such black holes may widely oscillate around the centre of the system  as they are only $\lesssim 10^4$ times heavier than the field population. Such wandering could in principle dynamically heat and puff up the system even more.

In our analysis, we further assumed that the UFDGs are spherically symmetric and in dynamical equilibrium. In fact, only a handful of them exhibit significant flattening and/or signs of rotation \citep{2012AJ....144....4M}. Tidal stirring and mass stripping from the primary halo could at first make both the stellar and dark matter distribution more elongated; however, dynamical relaxation is well known to make the system round over a sufficiently long time-scale even if the system was initially maximally flat, as was repeatedly shown by numerical simulations of both galaxy clusters \citep{2005MNRAS.364..607M} and  Milky Way-like halos \citep{2016ApJ...827L..15T,2018IAUS..334..209B}. The assumption of spherical symmetry made throughout this work thus appears reasonable. Concerning the dynamical state of the UFDGs, it is expected that phenomena which would put the UFDGs out of equilibrium, e.g. tidal stirring, tendentially lead to extra dynamical heating of the stars \citep{2010AdAst2010E..25M,2016ApJ...827L..15T}. Therefore, the stellar systems would puff up further and exhibit increased velocity dispersion and half-mass radii, thus making our findings on the evolution of both quantities conservative.

Having discussed some possible shortcomings of  our method, we highlight that our results join the class of dynamical constraints that severely challenge the PBH-DM scenario. The PBH mass window previously left open was already in conflict with other constraints \citep{Carr_et_al_2017}. Here, we closed this window by improving previously applied methods in several ways. In particular, we were also able to show that the increase of the stellar velocity and the dynamical friction due to PBHs actually lower the heating efficiency in a non-negligible way, making our results more precise than previous ones.

In light of these constraints, the PBH-DM scenario appears unlikely to be realised in our Universe, and the very nature of DM remains elusive.

\section*{Acknowledgements}
\JS{We thank Eugene Vasiliev and an anonymous referee for helpful advice and input}. Numerical simulations were performed on the ETH Euler cluster. We acknowledge \citet{Hinton2016} for the python package {\fontfamily{qcr}\selectfont ChainConsumer} that has been widely used for the production of most figures and tables. EB, PRC, and LM acknowledge support from the Swiss National Science Foundation under the Grant 200020\_178949.

\scalefont{0.94}
\setlength{\bibhang}{1.6em}
\setlength\labelwidth{0.0em}
\bibliographystyle{mnras}
\bibliography{improved_constraints_on_PBHs_as_DM}
\normalsize

\label{lastpage}
\end{document}